\documentclass[12pt,preprint]{aastex}

\newcommand{\Swift}{\textit {Swift} }

\newcommand{\akari}{\textit {AKARI} }

\newcommand{\cgs}{${\rm erg}\,{\rm cm}^{-2}\,{\rm s}^{-1}$\,}
\newcommand{\ph}{${\rm ph}\,{\rm cm}^{-2}\,{\rm s}^{-1}$\,}
\newcommand{\Rrx}{$R_{\rm rX}$ }
\newcommand{\Rro}{$R_{\rm rB}$ }
\newcommand{\degree}{$^{\circ}$}
\newcommand{\lat}{\textit {Fermi}--LAT }
\newcommand{\bat}{\textit {Swift}--BAT }

\slugcomment{Accepted for publication in The Astrophysical Journal}
\shorttitle{\lat Seyferts}
\shortauthors{\lat}

\begin{document}

\title{Search for Gamma-ray Emission from \\
 X-ray Selected Seyfert Galaxies with \lat}
\author{
M.~Ackermann\altaffilmark{2}, 
M.~Ajello\altaffilmark{2}, 
A.~Allafort\altaffilmark{2}, 
L.~Baldini\altaffilmark{3}, 
J.~Ballet\altaffilmark{4}, 
G.~Barbiellini\altaffilmark{5,6}, 
D.~Bastieri\altaffilmark{7,8}, 
K.~Bechtol\altaffilmark{2,1}, 
R.~Bellazzini\altaffilmark{3}, 
B.~Berenji\altaffilmark{2}
% ...
}
\altaffiltext{1}{Corresponding authors: M.~Hayashida, mahaya@slac.stanford.edu; {\L}.~Stawarz, stawarz@astro.isas.jaxa.jp; K.~Bechtol, bechtol@stanford.edu; G.~M.~Madejski, madejski@slac.stanford.edu.}
\altaffiltext{2}{W. W. Hansen Experimental Physics Laboratory, Kavli Institute for Particle Astrophysics and Cosmology, Department of Physics and SLAC National Accelerator Laboratory, Stanford University, Stanford, CA 94305, USA}
\altaffiltext{3}{Istituto Nazionale di Fisica Nucleare, Sezione di Pisa, I-56127 Pisa, Italy}
\altaffiltext{4}{Laboratoire AIM, CEA-IRFU/CNRS/Universit\'e Paris Diderot, Service d'Astrophysique, CEA Saclay, 91191 Gif sur Yvette, France}
\altaffiltext{5}{Istituto Nazionale di Fisica Nucleare, Sezione di Trieste, I-34127 Trieste, Italy}
\altaffiltext{6}{Dipartimento di Fisica, Universit\`a di Trieste, I-34127 Trieste, Italy}
\altaffiltext{7}{Istituto Nazionale di Fisica Nucleare, Sezione di Padova, I-35131 Padova, Italy}
\altaffiltext{8}{Dipartimento di Fisica ``G. Galilei", Universit\`a di Padova, I-35131 Padova, Italy}
% ...

\begin{abstract}
We report on a systematic investigation of the $\gamma$-ray properties of 120 hard X-ray--selected Seyfert galaxies classified as `radio-quiet' objects, utilizing the three-year accumulation of \lat data. Our sample of Seyfert galaxies is selected using the \bat 58-month catalog, restricting the analysis to the bright sources with average hard X-ray fluxes $F_{14-195\,{\rm keV}} \geq 2.5\times10^{-11}$\,\cgs at high Galactic latitudes ($|b|>10^{\circ}$). In order to remove `radio-loud' objects from the sample, we use the `hard X-ray radio loudness parameter', \Rrx, defined as the ratio of the total 1.4\, GHz radio to $14-195$ keV hard X-ray energy fluxes. Among 120 X-ray bright Seyfert galaxies with \Rrx\,$<10^{-4}$, we did not find a statistically significant $\gamma$-ray excess ($TS>25)$ positionally coincident with any target Seyferts, with possible exceptions of ESO~323--G077 and NGC~6814. The mean value of the 95\,\% confidence level  $\gamma$-ray upper limit for the integrated photon flux above 100 MeV from the analyzed Seyferts is $\simeq 4 \times 10^{-9}$\,\ph, and the upper limits derived for several objects reach $\simeq 1 \times 10^{-9}$\,\ph. Our results indicate that no prominent $\gamma$-ray emission component related to active galactic nucleus activity is present in the spectra of Seyferts around GeV energies. The \lat upper limits derived for our sample probe the ratio of $\gamma$-ray to X-ray luminosities $L_{\gamma} / L_{\rm X} < 0.1$, and even $<0.01$ in some cases. The obtained results impose novel constraints on the models for high energy radiation of `radio-quiet' Seyfert galaxies.
\end{abstract}

\keywords{accretion, accretion disks --- galaxies: active --- galaxies: Seyfert --- gamma rays: galaxies --- X-rays: galaxies}

\section{Introduction}

The all-sky observations of celestial objects by the Large 
Area Telescope~\citep[LAT:][]{LAT} aboard the {\it Fermi} 
Gamma-ray Space Telescope confirmed that, in addition 
to Gamma-ray Bursts, there are at least two more general 
classes of bright extragalactic sources of $\gamma$-rays~\citep{1FGL}.
One class comprises Active Galactic Nuclei (AGN) with powerful 
relativistic jets, including blazars, radio-loud Narrow-line 
Seyfert 1 galaxies (NLS1s), and radio galaxies, which produce 
beamed high-energy emission via inverse-Compton scattering 
of soft photon fields on ultra-relativistic jet electrons. 
The other class consists of nearby galaxies with prominent 
starburst systems, which produce diffuse, un-beamed $\gamma$-ray 
emission resulting from the interactions of cosmic-ray particles 
with the interstellar medium (ISM). An important question arises 
whether those 
are the only classes of extragalactic $\gamma$-ray sources. 
This question motivated us to search systematically for GeV 
emission from Seyfert galaxies using \textit{Fermi}--LAT. 
Seyfert galaxies constitute the most numerous class of 
AGN in the local Universe (local number density 
$\sim 10^{-4}$\,Mpc$^{-3}$), but at the same time 
lack, in general, ultrarelativistic collimated outflows 
or starburst regions \citep[e.g.,][]{ost89}. Seyferts, 
hosted by late-type galaxies, were originally identified 
in the optical regime by the presence of strong emission 
lines from highly ionized gas in their spectra~\citep{Sey43}. 
They are believed to harbor super-massive~\citep[$\mathcal{M}_{\rm BH} 
\sim 10^6-10^9 \, M_{\odot}$; see e.g.,][]{ho02} black holes 
in their galactic centers, and are powered by the infalling 
matter which forms accretion disks emitting intense optical/UV 
continuum radiation.

Seyfert galaxies are generally much weaker radio emitters than radio quasars or radio galaxies, but they are not `radio silent'. In addition to the diffuse radio continuum originating in the ISM of their late-type hosts, about half of the nearby Seyferts possess compact non-thermal radio cores \citep{ulv89,kuk95,hoR01}, which are often accompanied by arcsecond-scale jets and jet-like features \citep[e.g.,][]{mid04,gal06,lal11}. However, the radio cores and jets witnessed in Seyferts are generally very different from those observed in radio quasars or radio galaxies. In particular, radio cores in Seyfert galaxies are characterized by only modest brightness temperatures \citep[see, e.g.,][]{ulv01, ho08}, and in general, show no indication of relativistic beaming. 

Some Seyfert galaxies also display broad permitted emission lines in their spectra. Presence of such lines anti-correlates with the obscuration of unresolved cores in the optical--to--soft X-ray regime. This fact led to the idea that objects either possessing or lacking broad lines are intrinsically the same, differing only in the orientation of the central engine to the line of sight due to the selective absorption of the core emission by the anisotropically distributed circumnuclear dust \citep[the so-called `unification scheme';][]{ant93}. Hence broad-line Seyferts are called `unobscured' (or `type 1'), while Seyferts without broad emission lines in their spectra are called `obscured' (or `type 2'). One should note that there are several intermediate classes of Seyferts with respect to the nuclear obscuration \citep[type 1.2, 1.5, etc.][]{ost77}, as well as the objects which intrinsically lack broad emission lines \citep[NLS1s; e.g.,][]{pog00,fos11a}.

While optical information is needed for proper classification of an astrophysical source as an AGN, X-ray characteristics are equally important in understanding the physics of central engines in active galaxies. Seyfert galaxies are ubiquitous X-ray emitters \citep[e.g.,][]{hoX01,Ter03,cap06}, and the class is generally known to be particularly bright in the hard X-ray regime \citep[above $10$\,keV; see, e.g.,][]{tue08,bec09}. This hard X-ray emission, typically in a form of a power-law continuum (photon indices $\Gamma_{\rm X} \simeq 2$) cutting-off around a few hundred keV \citep{gon96,zdz00}, is well understood as being due to optical/UV disk emission reprocessed in the clumpy, hot, but predominantly thermal coronae of accretion disks \citep[see, e.g.,][for reviews]{pou98,zdz99}.

On the other hand, spectral properties of Seyfert galaxies in the $\gamma$-ray regime (and especially at high- and very-high-energy $\gamma$-rays, i.e. at GeV--TeV photon energy ranges) are basically unknown because of the limited sensitivity of past $\gamma$-ray instruments. The upper limits derived using observations by the the Imaging Compton Telescope \citep[COMPTEL:][]{comptel} onboard the Compton Gamma-Ray Observatory (CGRO) are consistent with no significant emission component around 1 MeV \citep{mai95,mai97}. Similarly, observations with the Energetic Gamma-Ray Experiment Telescope~\citep[EGRET:][]{egret} did not result in any detection of Seyfert galaxies (individually, or as a class by means of a stacking analysis) above $100$\,MeV \citep{Lin93,Cil04}. One might therefore conclude that, despite some expectations (see Section~\ref{sec:HEemission} below) and 
unlike jet dominated sources (blazars, radio-loud NLS1s, or nearby radio galaxies),
Seyferts are particularly `$\gamma$-ray quiet'. This issue can now be addressed more robustly using the \lat instrument, simply because of its unprecedented sensitivity to photons on the GeV range.

\lat has already discovered or confirmed a number of different 
classes and types of non-blazar $\gamma$-ray--emitting AGN, such as 
NLS1s \citep{NLS}, low-power FR\,I radio galaxies \citep{MAGN}, 
high-power broad-line radio galaxies \citep{kat11}, and sources 
hosting `reborn' compact radio structures \citep{mcc11}. 
All these targets appear however to posses relativistic jets 
aligned relatively closely to the line of sight. Nearby starburst 
systems have been detected by \lat as well \citep{1FGL,len10,SB}. 
However, radio-quiet Seyfert galaxies lacking a circumnuclear 
starburst have never been significantly detected as $\gamma$-ray sources. 
We note that, in parallel to our studies, \citet{ten11} have reported their analysis 
of 491 Seyfert galaxies included in the \bat catalog using 
2.1 years accumulation of \lat data in the 1--100 GeV energy range. 
\citet{ten11} found only two objects in their sample, NGC 1068 and NGC 4945, 
to be significantly detected in the 1--100 GeV energy range.
Those two sources have been already reported as $\gamma$-ray 
emitters in the First \lat Catalog~\citep{1FGL} and discussed 
in more detail by~\citet{len10}, but their GeV emission most likely 
originates in the ISM of the host galaxies~\citep{SB}.  
In this paper we report on a systematic and detailed investigation 
of the $\gamma$-ray properties of hard X-ray--selected Seyfert 
galaxies classified as radio-quiet objects, utilizing the 
three-year accumulation of the \lat data from 0.1--100~GeV, and
report flux limits for individual sources.
We also discuss the 
derived upper limits compared with fluxes in other wavebands for each source.
The paper is organized as 
follows: in \S\,2 we discuss the sample selection; the \lat data 
analysis and the results are presented in \S\,3 and \S\,4, 
respectively; the final discussion of our results are given in \S\,5. 

\section{Sample Selection}

Observations in hard X-rays are useful for selecting a complete and unbiased sample of Seyfert galaxies because hard X-ray emission is a clear and common signature of AGN activity, as described in the previous section. By contrast, the optical--to--soft X-ray emission of Seyfert galaxies may be subject to severe obscuration by circumnuclear dust, depending upon the orientation of the source to the line of sight. The Burst Alert Telescope~\citep[BAT:][]{bat} onboard the \Swift satellite has provided all-sky survey data in the hard X-ray band with unprecedented high sensitivity, which are well-suited for our investigation given the similar observational strategies of \bat and \lat. During the last five years of the \bat observations, about seven hundred AGN and galaxies were detected above $15$\,keV~\citep{Bau10,Cus10}. Notably, Seyferts outnumber the other classes of AGN detected in the hard X-ray band.

For this project we have selected a sample of the hard X-ray brightest Seyfert galaxies using the most recent version of the publicly available \bat 58-month catalog\footnote{\texttt{http://heasarc.gsfc.nasa.gov/docs/swift/results/bs58mon/}}, restricting the analysis to sources with average $14-195$\,keV fluxes equal to or greater than $2.5\times10^{-11}$\,\cgs. Such hard X-ray flux selection returns 179 non-blazar type AGN which are classified as either `galaxies' or `Seyfert galaxies' in the \bat 58-month catalog. From these, we excluded sources located close to the Galactic Plane, specifically those within Galactic latitudes $|b| < 10$\degree\ for the Galactic longitudes $|l|>20^{\circ}$, and $|b| < 20$\degree\ for $-20^{\circ} < l < 20^{\circ}$, because \lat sensitivity is reduced towards the Galactic plane due to substantial foreground emission related to the ISM of our Galaxy, and presence of numerous Galactic $\gamma$-ray emitters~\citep{1FGL}. All selected sources are also included in another independent \bat catalog: the Palermo \bat 54-month catalog~\citep{Cus10}.

The constructed sample is contaminated by several objects with bright relativistic jets such as nearby radio galaxies (e.g., Centaurus A) and radio-loud quasars, which can be classified also as Seyferts based on their emission line spectral properties in the optical band. All such sources should be removed from the analyzed sample, since those AGN are physically distinct from `classical' Seyferts. In principle, this could be accomplished by investigating the `radio loudness' parameter for the selected targets, i.e. the ratio between the monochromatic 5\,GHz radio and $B$-band optical fluxes, \Rro\,$\equiv F_{\rm 5\,GHz}/F_{\rm B}$. This parameter is often used to distinguish radio-loud (\Rro\,$>10$) from radio-quiet (\Rro\,$<10$) quasars, according to the criteria proposed by \citet{Kel89}, and is widely accepted as a useful proxy for the jet production efficiency. However, such an interpretation holds {\it only} if the radio fluxes correspond strictly to the jet emission, and the $B$-band optical fluxes are mainly due to the accretion disk emission. Both the total optical and radio fluxes in Seyferts can be dominated by host galaxies. If no careful subtraction of the starlight emission is performed, all Seyfert galaxies appear to be radio quiet (with \Rro\,$<10$). Yet when the starlight emission is carefully subtracted, many `classical' Seyfert galaxies (especially those accreting at lower rates) formally become radio loud, even if core radio fluxes are used instead of the total radio fluxes, as demonstrated first by \citet{ho01}, \citet{ho02}, and discussed further by \citet{sik07}. Another problem is that if one is dealing with a mixture of type 1 and type 2 Seyferts, the intrinsic nuclear optical fluxes may be extremely difficult to determine for the obscured (type 2) objects.  

For these reasons, we conclude that the standard definition of the radio loudness parameter is not well-suited for our purposes.  Instead, we use the `hard X-ray radio loudness parameter', $R_{\rm rX}$, defined as the dimensionless ratio of monochromatic radio (1.4\,GHz) energy flux density to integrated hard X-ray ($14-195$\,keV) energy flux density,
\begin{equation}
R_{\rm rX} = \frac{[\nu F_{\nu}]_{\rm 1.4\,GHz}}{F_{\rm 14-195\,keV}} \, .
\end{equation}
An analogous X-ray radio loudness parameter was first introduced for Seyfert galaxies and Low-Ionization Nuclear Emission-line Regions (LINERs) by \citet{Ter03}, and discussed further by \citet{pan07}. However, those authors used X-ray data from a lower (medium) photon energy range $2-10$\,keV, rather than the hard X-ray fluxes considered in this work. Our choice of using the hard X-ray fluxes from the \bat catalog has an advantage of minimizing the effect of a possible absorption of the X-ray emission in obscured (type 2) objects. At the same time, the typical X-ray photon indices of unobscured Seyfert galaxies $\Gamma_{\rm X} \lesssim 2$ within the medium range~\citep[e.g.,][claiming $\Gamma_{\rm X} \approx 1.74 \pm 0.02$ for the $2-10$\,keV band]{zho10} and $\Gamma_{\rm X} \gtrsim 2$ at hard X-rays~\citep[e.g.,][reporting $\Gamma_{\rm X} \approx 2.23 \pm 0.11$ in the $14-195$\,keV band]{Aje08}, imply roughly comparable \textit{intrinsic} energy flux densities in both X-ray regimes. Therefore, the radio loudness parameters evaluated using the definition introduced here and the definition of \citet{Ter03} or \citet{pan07}, should be roughly equivalent. On the other hand, in the case of very Compton-thick objects with an intrinsic absorption column density $N_{\rm H}$ of more than $10^{24.5}$\,cm$^{-2}$, even hard X-ray fluxes in the $14-195$ keV are affected by absorption~\citep[see e.g.,][]{gil07}; hence the radio loudness parameters provided for such sources have to be taken with caution.

In order to evaluate the radio loudness parameter \Rrx for all the analyzed objects, we gather their total radio fluxes from the literature including catalogs such as NRAO VLA Sky Survey~\citep[NVSS;][]{NVSS}, the VLA Faint Images of the Radio Sky at Twenty-cm~\citep[FIRST;][]{FIRST}, or Parkes Catalogue 1990~\citep[PKSCAT90;][see Table~\ref{tab:samples}]{PKS}. We use the $1.4$\,GHz fluxes, because the data in this band have much better coverage than at $5$\,GHz. In the case of sources for which $1.4$\,GHz fluxes are not available, we use measurements at other frequencies ($\nu=0.843$ or $4.86$\,GHz) \citep{SUMSS, Mil93}, and convert those fluxes $F_{\nu}$ to fluxes at $1.4$\,GHz as $[\nu F_{\nu}]_{\rm 1.4\,GHz} = (\nu_{\rm 1.4\,GHz}/\nu)^{1-\alpha} \, [\nu F_{\nu}]$ assuming a universal radio spectral index $\alpha = 0.7$ for non-blazar type AGN. We note that among the analyzed objects there are seven Seyfert galaxies for which radio data are not available in the literature; the radio loudness parameters for these cannot be thus evaluated.

\begin{figure}
\epsscale{.80}
\plotone{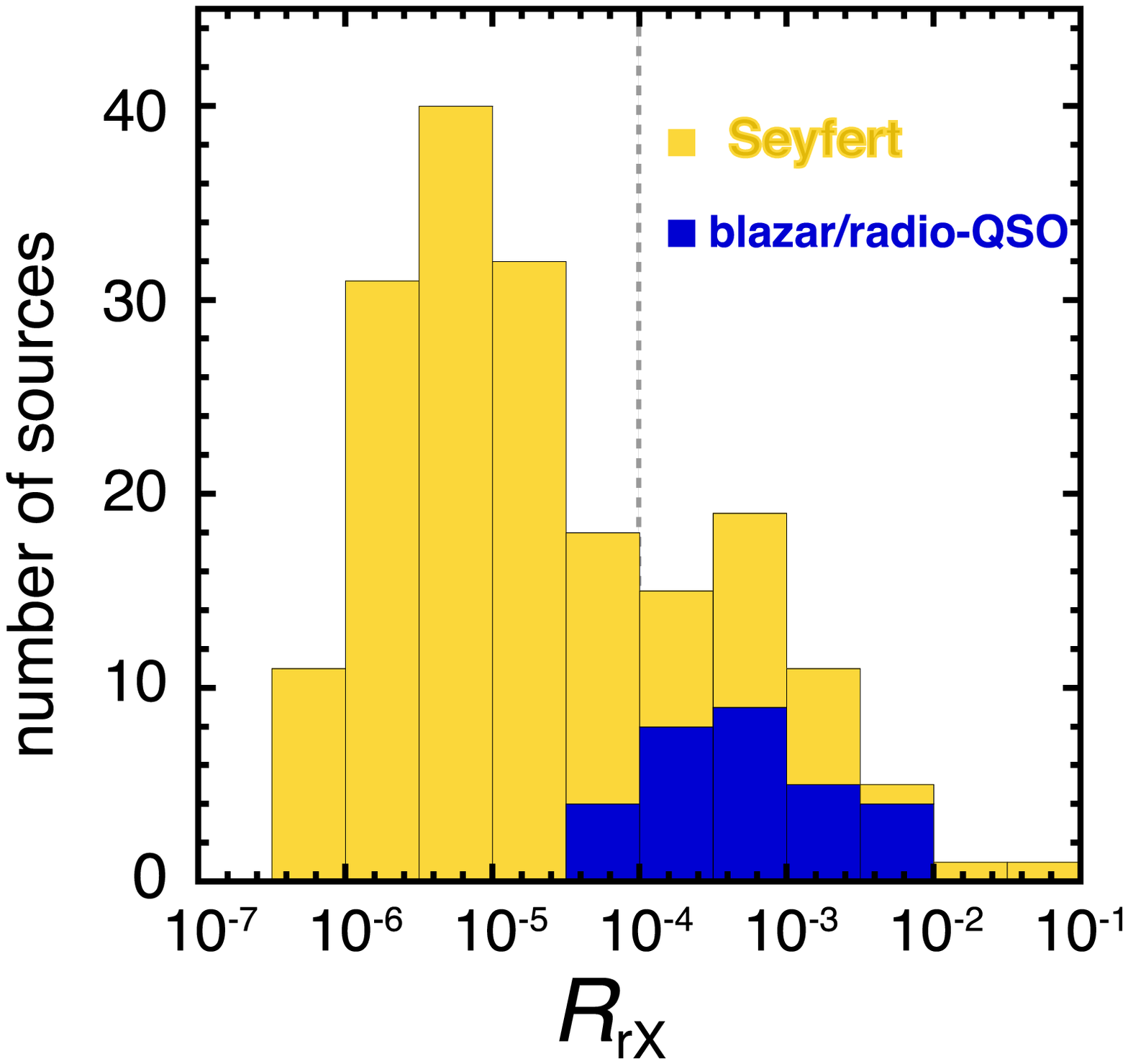}
\caption{Distribution of the `hard X-ray radio loudness parameter' \Rrx for Seyfert galaxies selected from the \bat 58-month catalog based on the flux and position cuts described in \S\,2 (yellow bars), and also for the comparison sample of blazars and radio quasars (blue bars).}
\label{fig:Rrx}
\end{figure}

Figure~\ref{fig:Rrx} shows a histogram of the \Rrx distribution for the Seyfert galaxies selected from the \bat 58-month catalog after the flux and position cuts described above (yellow bars in the figure). For comparison, we also plot the distribution of the \Rrx parameter derived for classical `radio-loud' AGN, which are dominated by the beamed emission of relativistic jets. These latter sources are similarly selected from the \bat 58-month catalog, based on the provided BAT classification (blazar or radio-quasar)\footnote{they are categorized as `beamed AGN' in the BAT catalog} and hard X-ray fluxes $\geq 2.5\times10^{-11}$\,\cgs. The selected blazars and radio-quasars (blue bars in Figure~\ref{fig:Rrx}) are characterized by higher values of the radio loudness parameter and different \Rrx distribution when compared to the analyzed population of Seyferts. 
As indicated by Figure~\ref{fig:Rrx}, the critical value \Rrx\,$= 10^{-4}$ may be used to differentiate between the truly radio-loud and radio-quiet objects, and this cut is applied in our analysis further below. 

\citet{Ter03} and \citet{pan07} found that \Rro\,$\sim 10^5$\,\Rrx for Seyferts and low-luminosity AGN using their medium X-ray fluxes. Since the intrinsic energy flux densities in the medium and hard X-ray regimes are expected to be comparable for Seyferts (see discussion above), the hard X-ray loudness parameter value \Rrx\,$= 10^{-4}$ roughly corresponds to the `classical' radio loudness parameter \Rro\,$= 10$.
This simple conversion may not be correct in all cases, though. In particular, in the comparison sample of blazars, there are four objects characterized by the `standard' radio-loudness parameters \Rro\,$> 10$ but \Rrx\,$< 10^{-4}$, namely 2MASS~J16561677--3302127, QSO~B0033+595, Mrk~421, and QSO~B0229+200. Three of them are well known `high-frequency peaked' BL Lac objects (HBLs) for which the X-ray fluxes are uniquely dominated by the synchrotron emission of highly relativistic jets, and as a result their X-ray--defined radio loudness parameters are low. Those are however exceptional objects in the \bat catalog. On the other hand, for some particularly low-luminosity spiral-hosted AGN (such as LINERs), the evaluated X-ray loudness parameters are \Rrx\,$>10^{-4}$, even though such sources lack relativistic jets. This is simply due to the fact that the total radio emission of such AGN is heavily dominated by the ISM, and is therefore relatively pronounced, while the total accretion-related X-ray emission is particularly low due to very low accretion rates in their nuclei. As a result, the evaluated X-ray loudness parameters for low-luminosity AGN accreting at low rates are high \citep[see in this context][]{Ter03,ho01}.

In order to check our final sample against contamination by objects containing prominent relativistic jets, first we check 12 sources with relatively high \Rrx values: $10^{-4.5} < R_{\rm rX} < 10^{-4.0}$. Among them, 8 sources are obscured Seyferts (type 1.8--2), for which relatively high values of \Rrx could result from the absorption of the X-ray continuum rather than prominent jet activity. The other 4 sources are type 1--1.5, namely Mrk~6, Mrk~1501, NGC~7469 and NGC~4051, for which no prominent relativistic jet is confirmed except for Mrk~1501. Only 4 sources (Mrk 1501, Mrk 348, NGC 3516 and NGC 7213) among our final sample of 120 sources have counterparts in the CRATES catalog~\citep{CRATES}, which provides a flux-limited all-sky survey of radio core emission. This suggests that most sources in our sample do not have a bright radio core and even the three CRATES Seyferts aside from Mrk~1501 do not display signatures of compact relativistic jets. Therefore, only Mrk~1501 in our sample shows peculiar features and, in fact, the source is known as a `radio-intermediate' source \citep[e.g.,][]{Mil93}. This galaxy is still worth including in our final sample to address a possible connection between `classical' radio-loud and radio-quiet AGN. We have thereby confirmed that our sample consists of `radio-quiet' Seyfert galaxies with a single peculiar `radio-intermediate' Seyfert object, Mrk 1501. 

Finally, we note that two starburst galaxies, NGC 1068 and NGC 4945, which are at the same time high accretion-rate Seyferts~\cite[e.g.,][]{lod03}, and which have been recently detected by \lat \citep{1FGL,len10,SB}, do not survive the applied cut in the radio loudness parameter, and therefore are not included in the analyzed sample. Both sources are however established Compton-thick objects, with nuclear hydrogen column densities $N_{\rm H} > 10^{24.5}$\,cm$^{-2}$ \citep[e.g.,][and references therein]{bur11}. As noted above, the hard X-ray fluxes of such objects are expected to be affected by nuclear obscuration, and as a result their X-ray radio loudness parameters may --- when uncorrected for the absorption --- formally read as \Rrx\,$> 10^{-4}$. Yet the GeV emission detected from those two sources most likely originates in the ISM of the galactic hosts, as discussed in detail in \citet{SB}, even though \citet{len10} claimed a dominant jet contribution for NGC~1068.

Summarizing, 120 sources are selected for the analysis accordingly to the following criteria:
\noindent
\begin{itemize}
\item hard X-ray fluxes $F_{\rm 14-195\,keV} \geq 2.5\times10^{-11}$\,\cgs in the \bat 58-month catalog;
\item spectral classification as `galaxies' or `Seyfert galaxies' in the \bat 58-month catalog;
\item hard X-ray radio loudness parameters \Rrx\,$< 10^{-4}$;
\item Galactic coordinates $|b| > 10$\degree\ for $|l|>20^{\circ}$, and $|b| >20$\degree\ for $-20^{\circ} < l < 20^{\circ}$.
\end{itemize}
\noindent
Table~\ref{tab:samples} provides source information for the constructed sample of objects including 62 Seyferts of type 1--1.5, 55 Seyferts of type 1.8--2, and three low-luminosity Seyferts classified as `galaxies' in the \bat catalog. 
The selected sample includes several radio-quiet NLS1s, such as NGC 4051, NGC 5506, and NGC 7314.
We emphasize once more that the applied cut in the hard X-ray--defined radio loudness parameter results in the rejection of not only truly radio-loud AGN, but also some Compton-thick Seyferts or low-luminosity low-accretion rate AGN. 

\section{\lat Data Analysis}\label{analysis}

\lat is a pair-production telescope with large effective area (6500\,cm$^2$ on axis for $>1$\,GeV photons) and large field of view (2.4\,sr at 1\,GeV), sensitive to $\gamma$ rays in the energy range from $20$\,MeV to $> 300$\,GeV. Full details of the instrument, as well as of the on-board and ground data processing, are provided in \citet{LAT}. Information regarding on-orbit calibration procedures is given in \citet{Calib}. \lat normally operates in a scanning `sky-survey' mode, which provides a full-sky coverage every two orbits (3 hours). For operational reasons, the standard rocking angle (defined as the angle between the zenith and the center of the LAT field of view) for survey mode was increased from 35\degree\ to 50\degree\ on 2009 September 3.

The data used in this work comprise three years of \lat observations
carried out between August 4, 2008 and August 5, 2011, corresponding
to the interval from 239557414 to 334195202 in Mission Elapsed Time
(MET).
We performed the analysis following the LAT standard analysis procedure\footnote{see details in http://fermi.gsfc.nasa.gov/ssc/data/analysis/} using
the LAT analysis software, \textit{ScienceTools v9r25v1}, 
together with the \textit{P7SOURCE\_V6} instrument response functions.
We discard events with zenith
angles $> 100^{\circ}$ and exclude time periods when the spacecraft
rocking angle relative to zenith exceeded 52\degree\ to avoid
contamination of $\gamma$ rays produced in the Earth's
atmosphere. Events are extracted within a $15^{\circ} \times
15^{\circ}$ region of interest (RoI) centered on the position of each
object in our sample (listed in Table~\ref{tab:result}). For our
analysis, we accept the events with estimated energies in the range
between $100$\,MeV and $100$\,GeV.

Gamma-ray fluxes and spectra are determined by performing a binned maximum likelihood fit of model parameters with \textit{gtlike} for events binned in direction and energy. The target objects themselves are modeled as point sources with simple power-law photon spectra $d\mathcal{F}/dE = N \times (E/E_0)^{-\Gamma}$. 
The background model applied here includes standard models for the isotropic and Galactic diffuse emission components\footnote{{\it `iso\_p7v6source.txt'} and {\it `gal\_2yearp7v6\_v0.fits'}}.
In addition, the model includes point sources representing all $\gamma$-ray emitters within each RoI based on the Second \lat
Catalog~\cite[2FGL:][]{2FGL}. 
We examine the significances of $\gamma$-ray signals for the analyzed sources by means of their test statistic ($TS$) values based on the likelihood ratio test \citep{ML}. If no significant $\gamma$-ray excess above background is detected, we derive a $95\,\%$ confidence level (CL) upper limit for the integrated photon flux above 100\,MeV
$\mathcal{F}(>{\rm 100\,MeV}) = \int_{\rm 100\,MeV} dE \, (d \mathcal{F}/dE)$, 
using the Bayesian method~\citep{hel83} with a fixed photon index $\Gamma$. 
Here, we assume two values for the photon index: $\Gamma=2.5$, corresponding to the average photon index for the flux-limited sample of `flat-spectrum radio quasars' included in the 1st \lat AGN catalog \citep[1LAC;][]{1LAC}, and $\Gamma=2.2$, corresponding to the typical $\gamma$-ray photon index of LAT-detected starburst galaxies~\citep{SB}.
These values should be considered as examples only because spectral properties of Seyfert galaxies around GeV photon energies are unknown.

\section{Results}\label{results}

Our analysis results are summarized in Table~\ref{tab:result}. We require that two conditions be met in order to claim the detection of $\gamma$-ray emission from a target AGN. First, a significant $\gamma$-ray excess above backgrounds with $TS>25$ must be present at the location of the Seyfert as given in the table.\footnote{$TS=25$ with 2 degrees of freedom corresponds to an estimated $\sim4.6\, \sigma$ pre-trials statistical significance assuming that the null-hypothesis $TS$ distribution follows a $\chi^2$ distribution \citep[see][]{ML}.}  Second, we require a positional coincidence defined here as a target AGN existing within the 95\,\% confidence localization region of the $\gamma$-ray excess. Following these criteria, we did not find any significant $\gamma$-ray detections among the 120 Seyfert galaxies in our sample, with possible exceptions of ESO~323--G077 and NGC~6814.

$TS$ values above 25 were obtained at the optically-determined locations of ESO~323--G077 and NGC~6814.
The $\gamma$-ray source 2FGL~J1306.9--4028 
has been associated with ESO~323--G077 with a probability of 0.8, 
and 2FGL J1942.5--1024 has been associated with NGC~6814 with 0.91 probability 
according to the 2FGL catalog and the Second LAT AGN Catalog (2LAC)~\citep{2FGL, 2LAC}.
 The 2FGL catalog warns, however, that 
 ``{\it we expect up to $\sim2$ false positives among the Seyfert galaxy
  associations\footnote{The 2FGL catalog uses 27651 
  Seyfert galaxies in its automatic source association pipeline.} (cf.~Table 8).}'' 
We consider two possibilities in this work.
First, we analyze the RoI under the assumption that ESO~323--G077/NGC~6814 is detected by the LAT as 2FGL~J1306.9--4028/2FGL J1942.5--1024. Second, we consider the case in which the proposed associations are actually the result of chance spatial coincidences. In the second case, we compute a flux upper limit at the position of ESO~323--G077/NGC~6814 with 2FGL~J1306.9--4028/2FGL J1942.5--1024 included as a background source in the model for the RoI. 
The proposed associations may be reinforced by more a precise localization given additional exposure, or confirmed by the identification of correlated variability with another waveband. 
We confirm no significant variability both for 2FGL~J1306.9--4028 and 2FGL J1942.5--1024 during the observation period, and find that the spectral shapes are consistent with a simple power law.
No blazar in the Roma-BZCAT catalog \citep{BZCAT} nor any flat-spectrum radio source in the CRATES catalog \citep{CRATES} can be found within 0\degree.4 of 2FGL~J1306.9--4028 and 2FGL J1942.5--1024.

\begin{figure}
  \centering
  \plotone{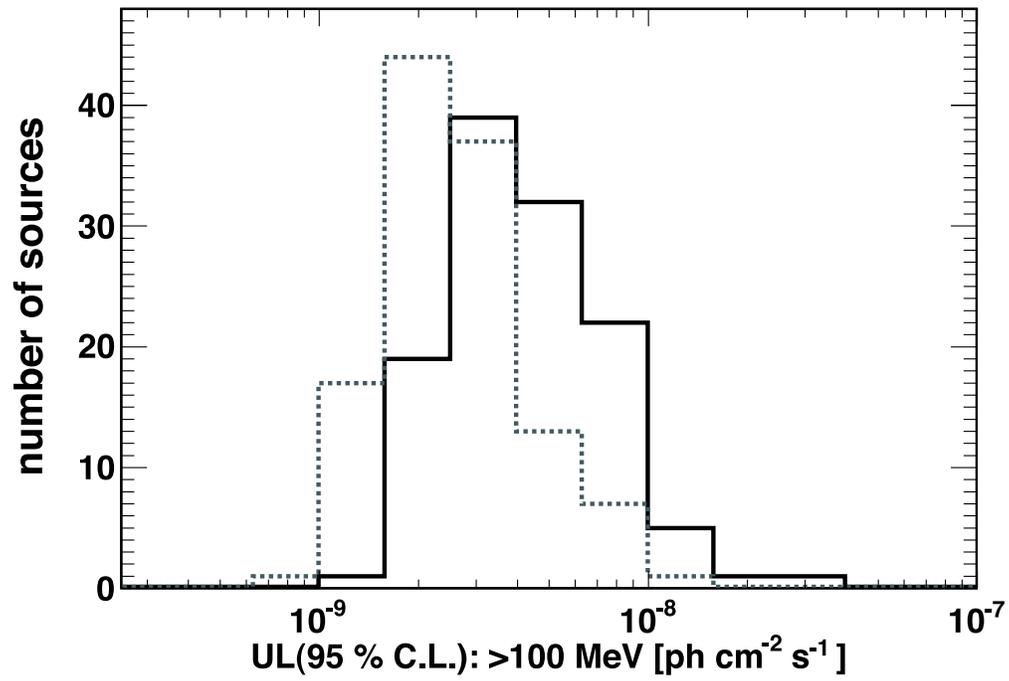}
\caption{Distribution of the \lat photon flux upper limits (95\,\% C.L.) for the analyzed sample of Seyfert galaxies calculated, assuming photon indices $\Gamma = 2.5$ (solid line) and $\Gamma = 2.2$ (dotted line) .
}
\label{fig:ULhist}
\end{figure}

Figure~\ref{fig:ULhist} shows the distribution of resulting upper 
limits for integrated photon fluxes above 100 MeV, 
$\mathcal{F}(>{\rm 100\,MeV})$. For instance, when we assume a photon 
index of 2.5, the mean value of the $\gamma$-ray upper limit from 
the analyzed Seyferts is $\simeq 4 \times 10^{-9}$\,\ph, 
and the upper limits derived for several objects are as low 
as $\simeq 1 \times 10^{-9}$\,\ph. The mean upper limit found with 
\lat data is therefore more than two orders of magnitude lower 
than the upper limits derived for the brightest Seyferts based on 
the \textit{SAS\,2} and \textit{COS\,B} data 
\citep[respectively]{big79,pol81}, more than an order of magnitude 
lower than the analogous EGRET upper limits, 
$(0.5-1.5)\times10^{-7}$\,\ph~\citep{Lin93}, and close to the 
lower bound of the effective upper limits from the EGRET stacking 
data analysis for the brightest 32 Seyfert objects, 
$(0.3-1.5)\times10^{-8}$\,\ph \citep{Cil04}.
We note here that \citet{ten11} estimated a typical flux 
upper limit of $\sim 1\times 10^{-10}$\, \ph above 1 GeV 
for a single source. 
This is consistent with our results 
covering the bandpass 0.1--100~GeV when 
re-scaling our $\gamma$-ray upper limits to their bandpass, 
assuming a power law spectral model with photon index 2.5.

\section{Discussion}

\subsection{Multiwavelength Comparison}

\begin{figure}
\epsscale{1.1}
\plottwo{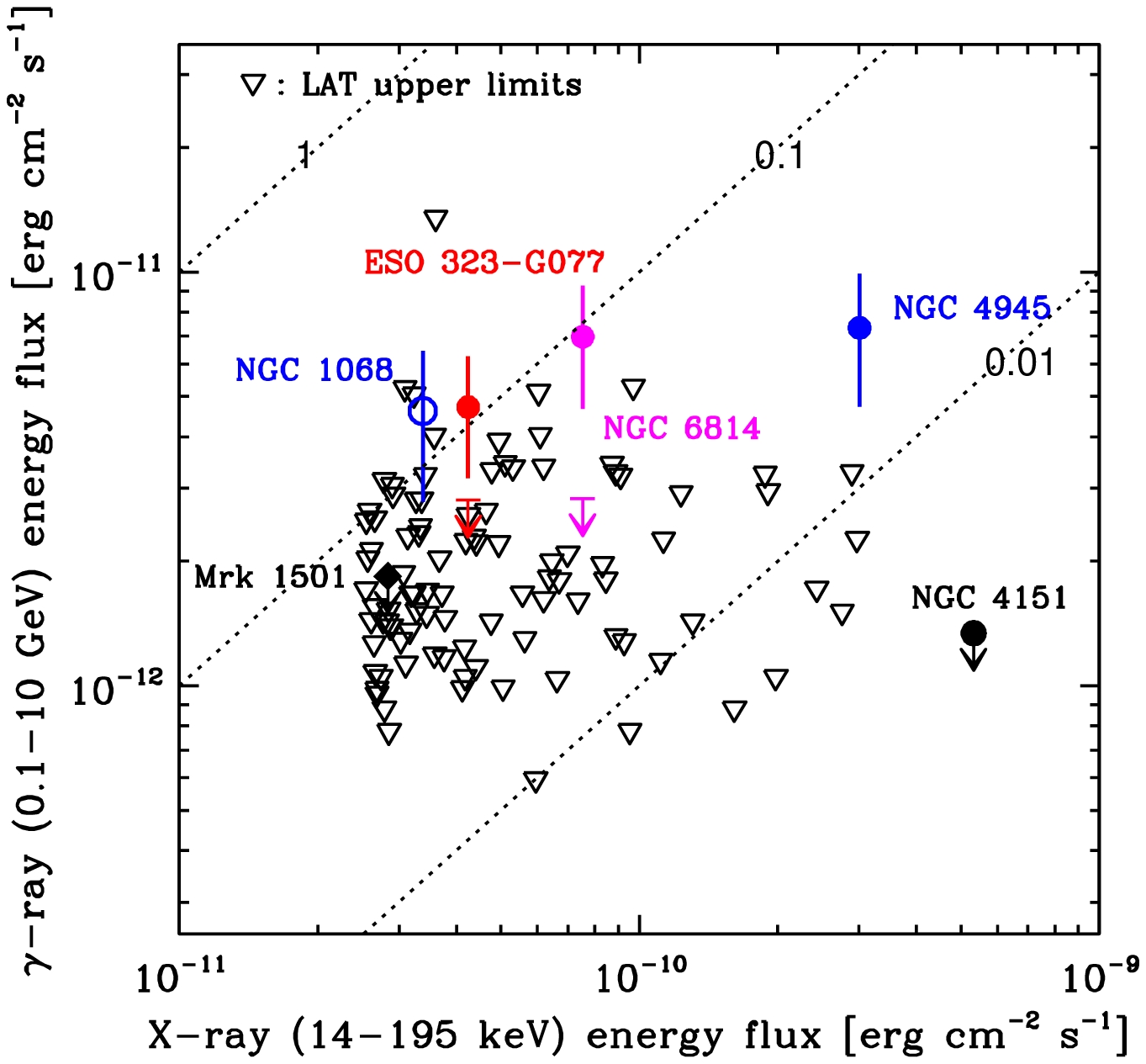}{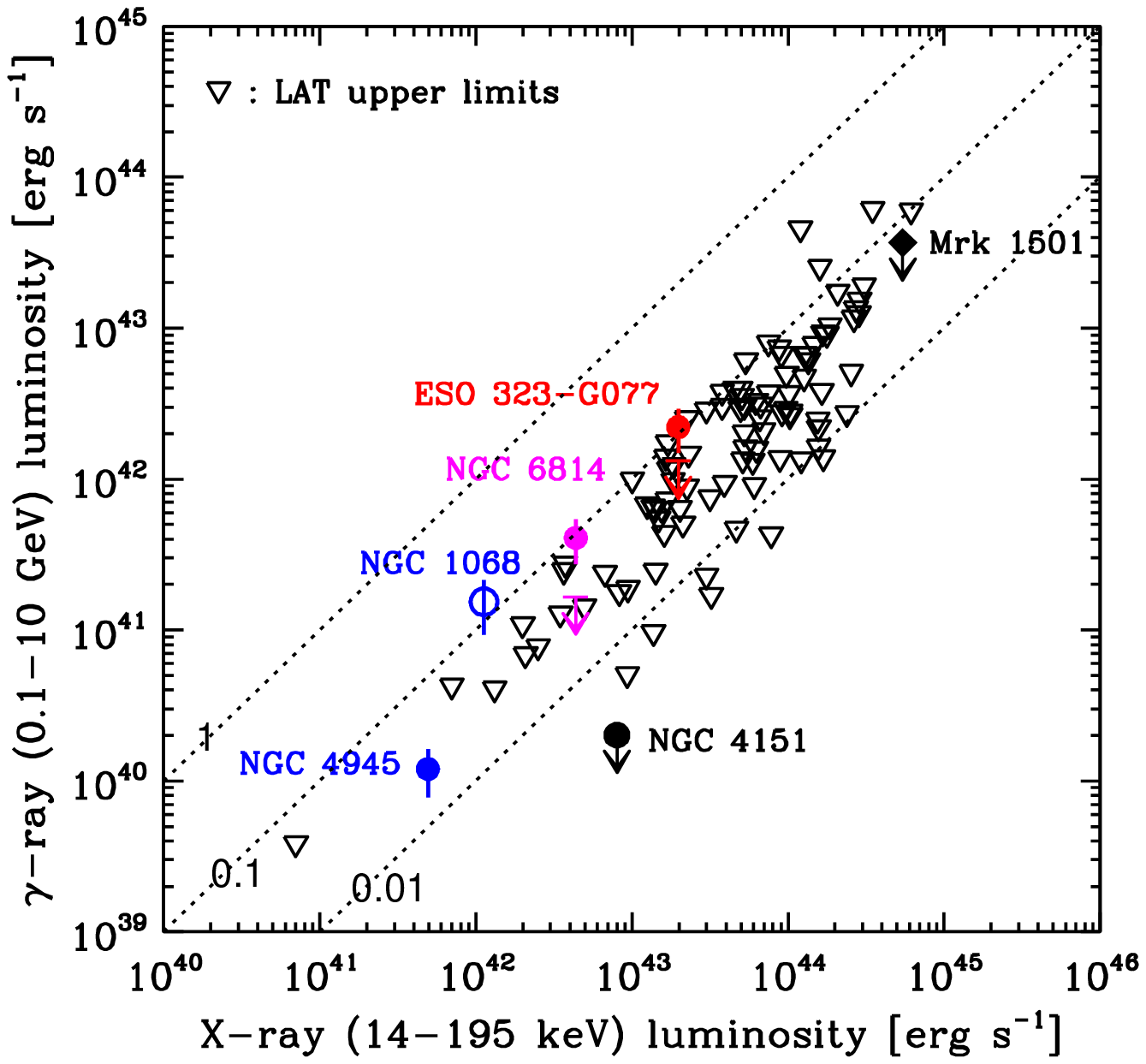}
\caption{{\it Left:} Hard X-ray ($14-195$\,keV) energy fluxes versus upper limits for the $\gamma$-ray ($0.1-10$\,GeV) energy fluxes for the analyzed sample of Seyferts (denoted by black open triangles) assuming a photon index $\Gamma=2.5$. Dotted lines from top left to bottom right denote the ratios between the $\gamma$-ray and hard X-ray energy fluxes 1, 0.1, and 0.01, respectively. 
Arrows denote ESO~323--G077 (red) and NGC~6814 (magenta) when the \lat upper limit is considered, and each flux is denoted by a filled circle
when assuming the associations with 2FGL~J1306.9--4028 and 2FGL J1942.5--1024, respectively.
The radio-intermediate quasar Mrk~1501 is denoted by a black filled diamond. NGC~4151 is marked by a black filled circle. For comparison, starburst galaxies NGC~1068 (blue open circle) and NGC~4945 (blue filled circle) are included. {\it Right:} The corresponding luminosity-luminosity plot.}
\label{fig:X_gamma}
\end{figure}

\begin{figure}
\epsscale{.80}
\plotone{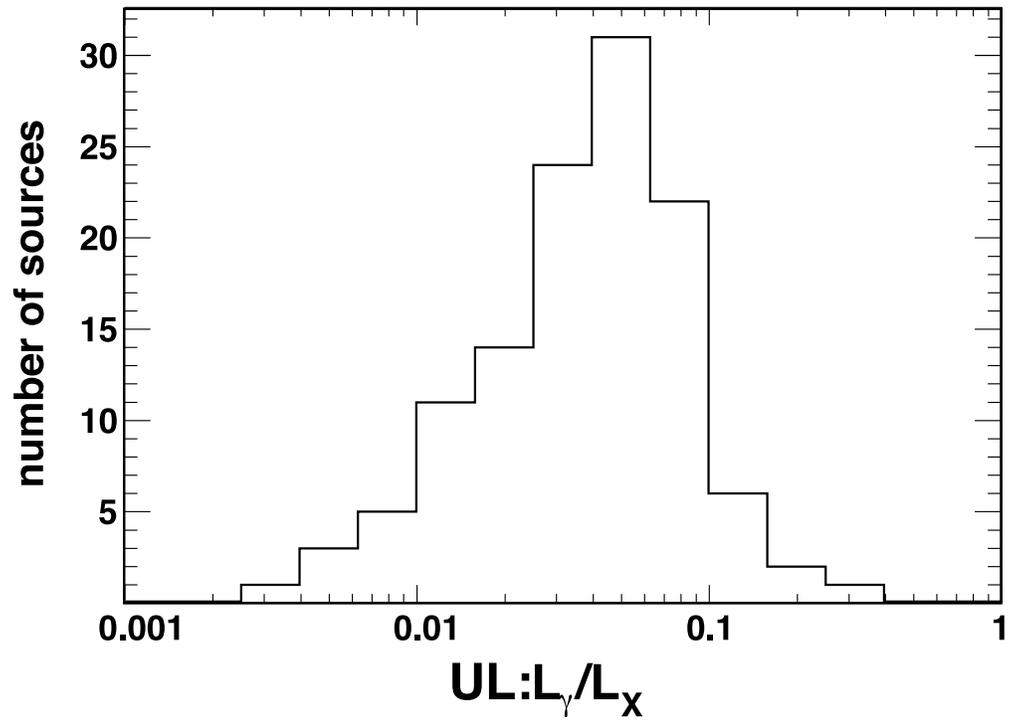}
\caption{Distribution of $\gamma$-ray-to-hard X-ray luminosity ratio for the analyzed sample of Seyfert galaxies, based on the $\gamma$-ray upper limit with an assumed photon index $\Gamma = 2.5$.}
\label{fig:histLgLx}
\end{figure}

The left panel of Figure~\ref{fig:X_gamma} compares hard X-ray ($14-195$\,keV) energy fluxes to upper limits for the $\gamma$-ray ($0.1-10$\,GeV) energy fluxes\footnote{Upper limits of the energy fluxes (and corresponding luminosities) are calculated with the upper energy bound of 10\, GeV based on the integrated photon flux upper limits above 0.1\,GeV.} for the analyzed sample of Seyfert galaxies (denoted in the figure by black open circles), together with a corresponding luminosity-luminosity ($L_{\gamma} - L_{\rm X}$) plot in the right panel. 
We discuss the $\gamma$-ray results based on the LAT upper limits derived with an assumed photon index of 2.5.
Dotted lines in the figures from top left to bottom right denote the ratios between the $\gamma$-ray and hard X-ray energy fluxes (or luminosities) $F_{\rm 0.1-10\,GeV}/F_{\rm 14-195\,keV} = 1$, 0.1, and 0.01, respectively. The distribution of the ratio between the $\gamma$-ray and hard X-ray luminosities is reported in Figure~\ref{fig:histLgLx}. As shown in the figure, for most of the analyzed objects the upper limits for this ratio are below 10\,\%, and for several particular sources are even below 1\,\%. The main conclusion here is that our investigation of the \lat data indicate that there is no emission component around GeV photon energies in Seyfert objects down to the level of $L_{\gamma}/L_{\rm X} < 0.1$ in most cases.

It is instructive to locate intriguing targets from the sample in the
parameter space of both panels of Figure~\ref{fig:X_gamma}, namely
ESO~323--G077 and NGC~6814, the radio-intermediate quasar
Mrk~1501, and the brightest hard-X-ray Seyfert galaxy in the sample
NGC~4151. For comparison, two LAT-detected starburst galaxies showing
Seyfert activity (NGC~1068 and NGC~4945) are plotted in the figure as
well. The multifrequency data together with the \lat fluxes for these
are taken from \citet[see also Table~\ref{tab:starburst}]{SB}. As
shown, although NGC~4151 is the brightest hard X-ray source among the
analyzed Seyfert galaxies, its intrinsic hard X-ray luminosity is
relatively modest, $L_{\rm X} \sim
10^{43}$\,erg\,s$^{-1}$. Importantly, the $\gamma$-ray--to--hard X-ray
luminosity ratio for this Seyfert is the lowest among our sample, 
$L_{\gamma}/L_{\rm X} \sim 0.0025$. This can be compared with
ESO~323--G077 and NGC~6814, for which the X-ray luminosities in
the BAT range are comparable to that of NGC~4151, but for which the
luminosity ratios would be $L_{\gamma}/L_{\rm X} \sim 0.11$ and
  $0.093$ in the case of the associations with 2FGL~J1306.9--4028 and
  2FGL~J1942.5--1024, respectively. Mrk~1501 is yet a different case, being characterized by a relatively low X-ray flux but high X-ray luminosity, $L_{\rm X} \gtrsim 10^{44}$\,erg\,s$^{-1}$. Indeed, this is the most distant object in the compiled sample. 
  The two starburst galaxies included here for comparison, NGC~4945 and NGC~1068, are characterized by low hard X-ray luminosities, $L_{\rm X} \sim 10^{42}$\,erg\,s$^{-1}$, and $\gamma$-ray--to--hard X-ray luminosity ratios $L_{\gamma}/L_{\rm X} \sim 0.1$.

\begin{figure}
\epsscale{1.1}
\plottwo{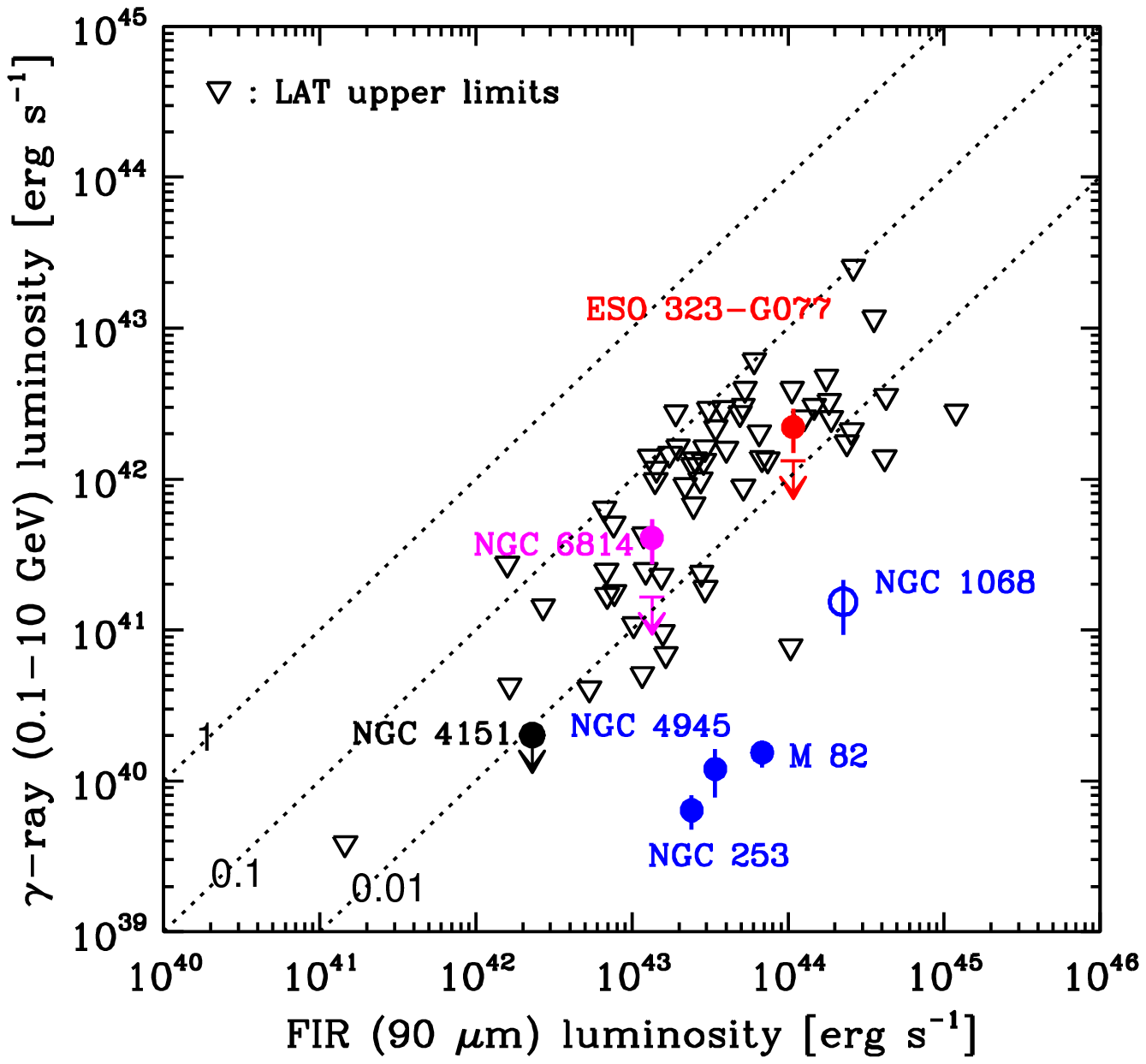}{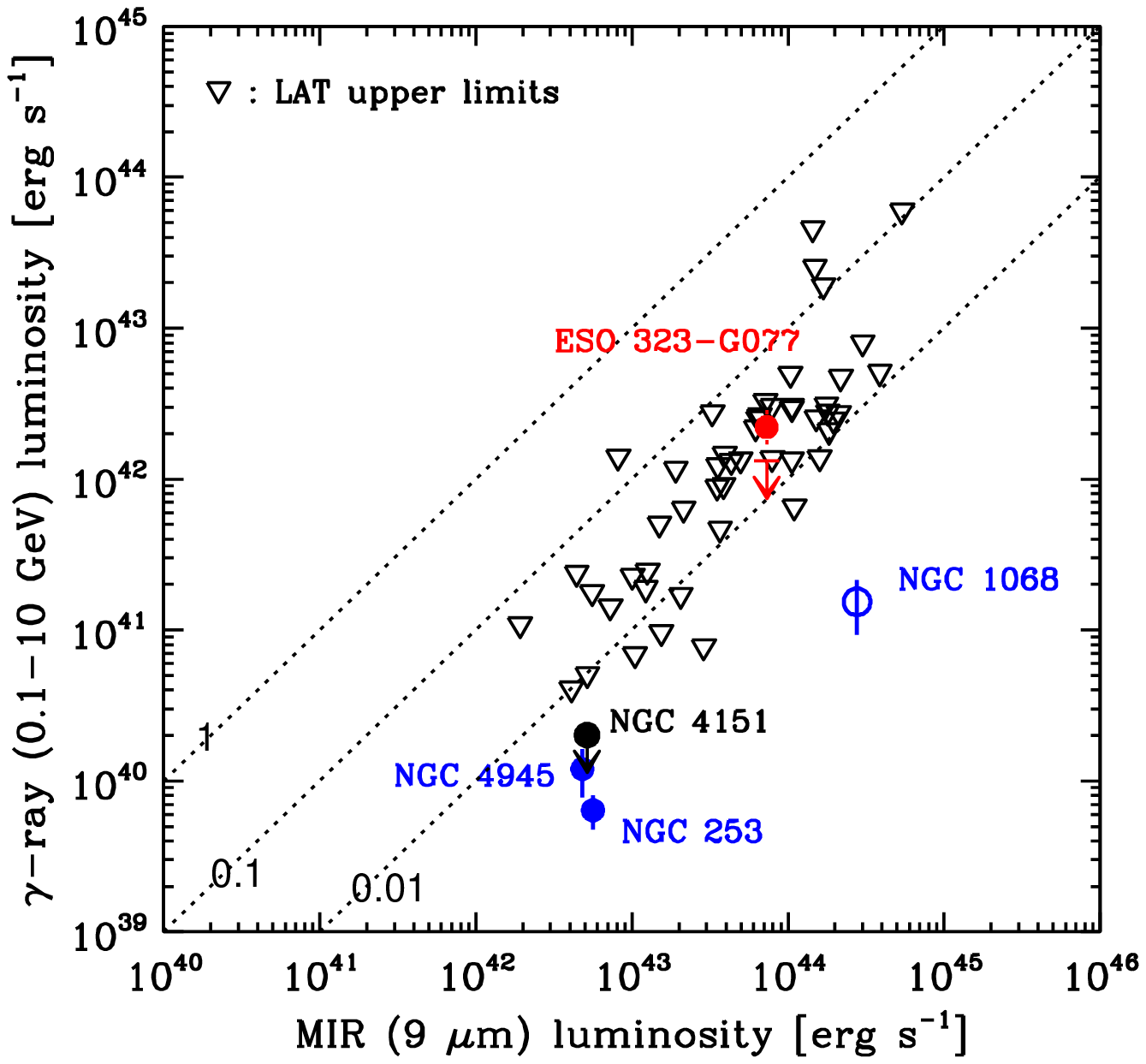}
\caption{Far-infrared ($90$\,\micron ; {\it left panel}) and mid-infrared ($9$\,\micron ; {\it right panel}) luminosities versus upper limits for the $\gamma$-ray ($0.1-10$\,GeV) luminosities for the analyzed sample of Seyferts assuming an photon index $\Gamma=2.5$. Dotted lines from top left to bottom right denote the ratios between the $\gamma$-ray and infrared luminosities 1, 0.1, and 0.01, respectively. 
Arrows denote ESO~323--G077 (red) and NGC~6814 (magenta) when the \lat upper limit is considered, and each flux is denoted by a filled circle
when assuming the associations with 2FGL~J1306.9--4028 and 2FGL J1942.5--1024, respectively.
No \akari 9\,\micron\ flux is available for NGC~6814.
NGC~4151 is marked by black filled circle. For comparison, starburst galaxies NGC~1068 (blue open circles), NGC~4945, NGC~253 and M~82 (blue filled circles) are included.}
\label{fig:IR_gamma}
\end{figure}

\lat upper limits derived for the analyzed Seyferts can be also compared with infrared fluxes measured by the \akari satellite. Here we use \akari 9\,\micron\ data \citep{AKARI-IRC} and 90\,\micron\ data \citep{AKARI-FIS} with a `good' quality ($FQUAL=3$), which are available for 65 and 73 sources from the analyzed sample, respectively. In the left and right panels of Figure~\ref{fig:IR_gamma} we present the corresponding luminosity-luminosity plots, including for comparison, the LAT-detected starburst galaxies NGC~1068, NGC~4945, NGC~253 and M~82 utilizing the \lat data analysis presented in \citet[see also Table~\ref{tab:starburst} below]{SB}. However, \akari 90\,\micron\ data for NGC~4945, NGC~253 and M~82 are flagged as `bad' quality ($FQUAL=1$). Hence, for these sources we use instead IRAS 60\,\micron\ data \citep{IRAS}, which can be considered as comparable to the \akari 90\,\micron\ data according to \citet{AKARI-FIS}. 

The far-infrared fluxes of Seyferts galaxies (`FIR'; 90\,\micron\ data) are expected to be dominated by thermal dust emission related to the star-forming activity of the galactic hosts, while mid-infrared fluxes (`MIR'; 9\,\micron\ data) may originate substantially from circumnuclear dust heated by accretion disk emission, i.e.~AGN activity. For most of the analyzed Seyferts the upper limits for the $L_{\gamma} / L_{\rm FIR}$ and $L_{\gamma} / L_{\rm MIR}$ ratios are in the range $0.01 - 0.1$ (cf. black dotted lines in the plots). At the same time, the LAT-detected starburst galaxies are characterized by $L_{\gamma} / L_{\rm FIR} \lesssim 0.001$. This suggests that detection of ISM emission in the GeV photon energy range from the bulk of the hard X-ray selected Seyfert objects --- emission analogous to that observed in nearby star-forming galaxies --- would require increasing the sensitivity of the \lat survey by roughly an order of magnitude. Yet, in the analyzed sample there are also some outliers with particularly high FIR luminosities and \lat upper limits low enough to already probe the GeV fluxes close to the expected level of the ISM-related $\gamma$-ray emission.

\begin{figure}
\epsscale{.80}
\plotone{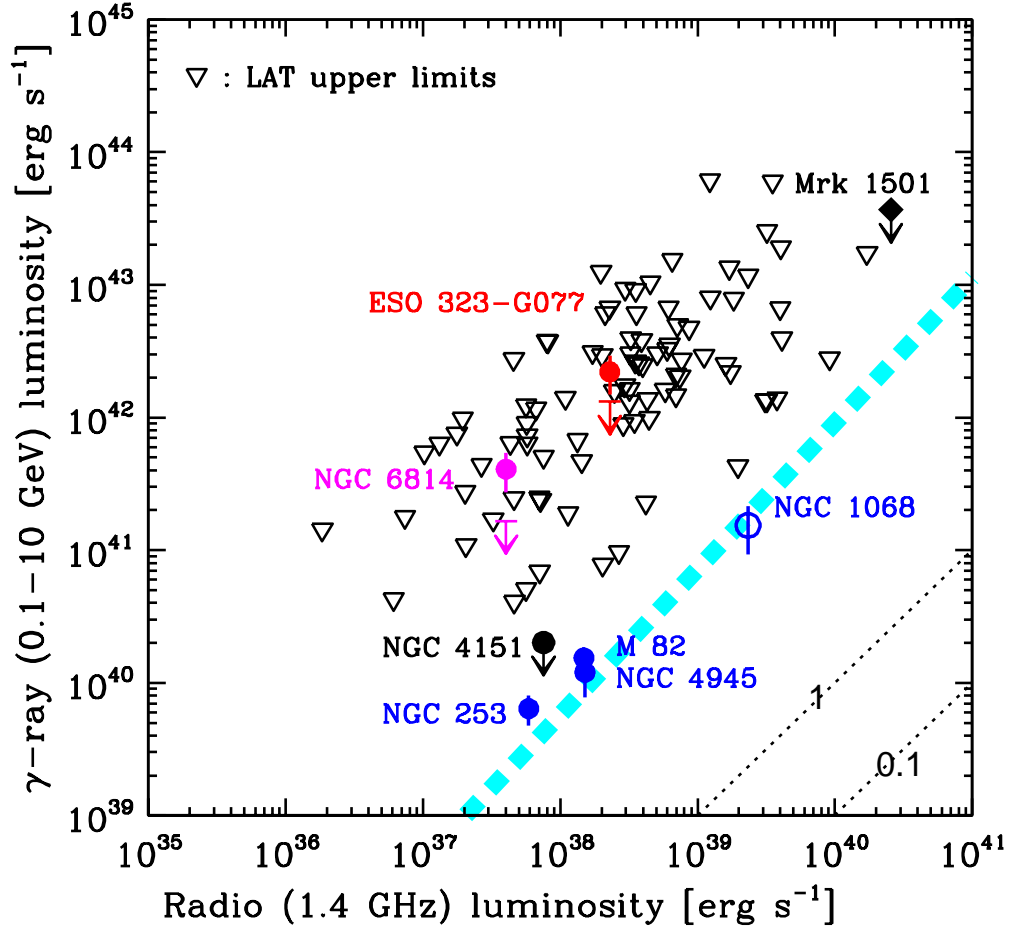}
\caption{Radio ($1.4$\,GHz) luminosities versus upper limits for the $\gamma$-ray ($0.1-10$\,GeV) luminosities for the analyzed sample of Seyferts assuming a photon index $\Gamma=2.5$. Black dotted lines from top left to bottom right denote the ratios between the $\gamma$-ray and radio luminosities 1 and 0.1, respectively. 
Arrows denote ESO~323--G077 (red) and NGC~6814 (magenta) when the \lat upper limit is considered, and each flux is denoted by a filled circle
when assuming the associations with 2FGL~J1306.9--4028 and 2FGL J1942.5--1024, respectively. 
The radio-intermediate quasar Mrk~1501 is denoted by a black filled diamond. NGC~4151 is marked by a black filled circle. For comparison, starburst galaxies NGC~1068 (blue open circle), NGC~4945, NGC~253 and M~82 (blue filled circles) are included. The thick dotted cyan line represents the best-fit power-law relation between the radio and GeV luminosities for star-forming and local galaxies discussed in \citet{SB}.}
\label{fig:Lr_gamma}
\end{figure}

A similar conclusion can be drawn from Figure~\ref{fig:Lr_gamma}, where we plot radio $1.4$\,GHz luminosities versus upper limits for the $\gamma$-ray luminosities derived for the analyzed Seyferts, including also the comparison sample of starburst galaxies. The thick dotted cyan line shown in the figure represents the best-fit power-law relation between the radio and GeV luminosities for star-forming and local galaxies discussed in \citet{SB}. The upper limits for the ratio $L_{\gamma} / L_{\rm R}$ in Seyfert sources are on average more than an order of magnitude above the $\gamma$-ray--to--radio luminosity ratios characterizing nearby star-forming galaxies. 
However, the relation between GeV and radio fluxes may not be expected to follow the trend established in star-forming systems for Seyferts 
in which AGN jet activity could contribute a substantial fraction of the total observed radio flux.
\citet{kat11} argued that the jet-related $\gamma$-ray emission of Seyfert galaxies is expected to be below the flux levels probed at present by \textit{Fermi}--LAT, at least for the majority of sources, and that conclusion is consistent with the upper limits presented in this work: $\gamma$-ray--to--radio luminosity ratio $L_{\gamma} / L_{\rm R} < 10^4$, on average. 

\begin{figure}
\epsscale{.80}
\plotone{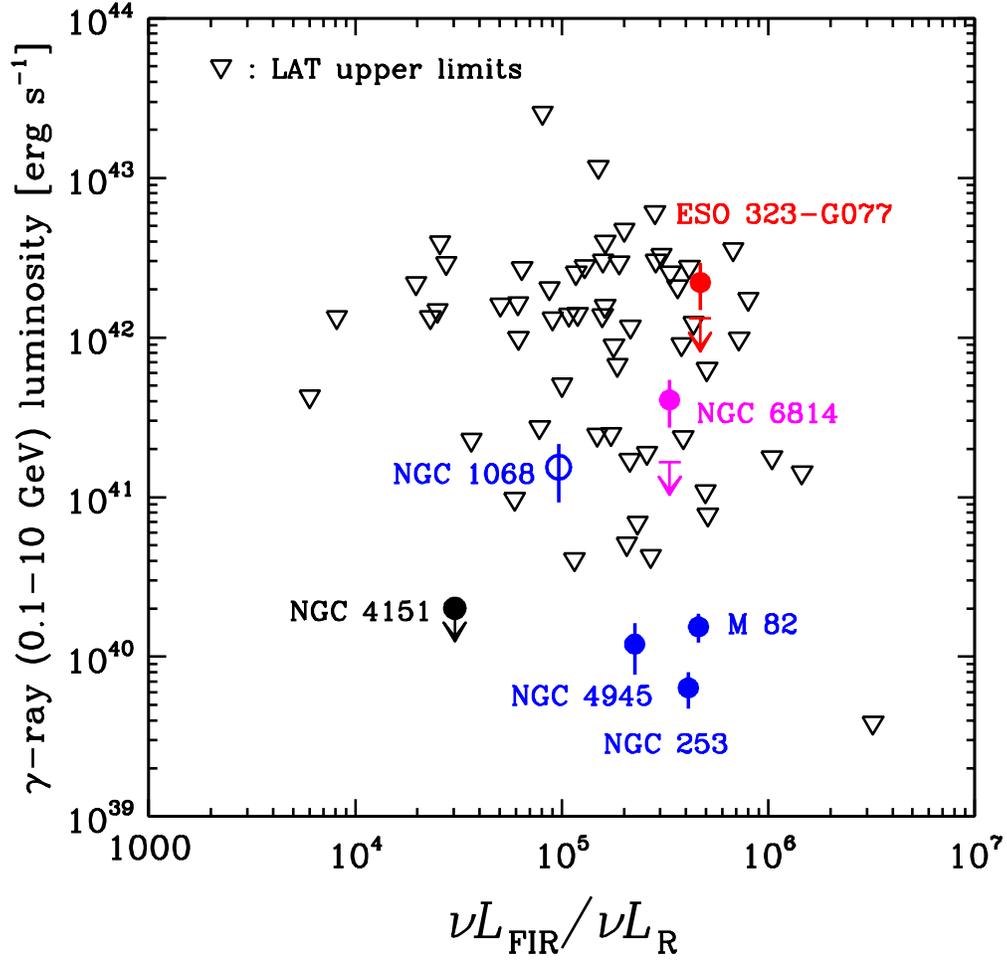}
\caption{Far-infrared--to--radio luminosity ratios versus upper limits for the $\gamma$-ray ($0.1-10$\,GeV) luminosities for the analyzed sample of Seyferts assuming a photon index $\Gamma=2.5$. 
Arrows denote ESO~323--G077 (red) and NGC~6814 (magenta) when the \lat upper limit is considered, and each flux is denoted by a filled circle
when assuming the associations with 2FGL~J1306.9--4028 and 2FGL J1942.5--1024, respectively.
NGC~4151 is marked by a black filled circle. For comparison, starburst galaxies NGC~1068 (blue open circle), NGC~4945, NGC~253 and M~82 (blue filled circles) are included.}
\label{fig:F90Fr_gamma}
\end{figure}

The issue of excess radio emission related to the jet activity in
Seyfert galaxies may be addressed by looking at the ratio of FIR and
radio luminosities for the considered targets, since a relatively
tight FIR--radio correlation has been established for non-active (and
therefore not jetted) galaxies \citep[see, e.g.,][]{yun01}. Seyferts
with particularly low FIR--to--radio luminosity ratios are likely
characterized by prominent jet activity. In
Figure~\ref{fig:F90Fr_gamma} we plot $L_{\rm FIR} / L_{\rm R}$ versus
upper limits for the $\gamma$-ray luminosities for the analyzed
sample. As expected, there are many hard X-ray--selected Seyferts
which are characterized by much lower $L_{\rm FIR} / L_{\rm R}$ ratios
than those in nearby starburst galaxies. For instance, NGC~4151, which
has a relatively prominent pc-scale and kpc-scale jet
\citep{mun03,ulv05}, shows one of the lowest $L_{\rm FIR} / L_{\rm R}$
ratio among the samples. In the case of ESO~323--G077 and
  NGC~6814 --- which are not characterized by any outstanding
radio or infrared luminosity when compared with the other Seyferts
included in the sample --- the ratio of the FIR and radio luminosities
is very similar to that observed in the LAT-detected starburst
galaxies, implying that ESO~323--G077 and NGC~6814 obey the
FIR-radio correlation well, and hence that there is not much room for
jet activity in these sources. This is in agreement with a
non-detection of a compact radio core in ESO~323--G077 by
high-resolution VLBI radio observations \citep{cor03}, 
and with the presence of only weak steep-spectrum 
radio core in NGC 6814~\citep{ulv84}.

\subsection{Possible $\gamma$-ray Emission Components in Seyfert Galaxies}
\label{sec:HEemission}

The \lat upper limits derived for the Seyfert galaxies in our sample probe the $\gamma$-ray luminosity range $L_{\gamma} / L_{\rm X} < 0.1$, and even $<0.01$ in some cases. Since hard X-ray luminosity is expected to constitute about 10\,\% of the bolometric AGN-related luminosity of a typical Seyfert galaxy \citep[see][]{ho08}, the results indicate that there is no emission component in Seyfert spectra at GeV photon energies down to the level of $1\,\%$ of the bolometric AGN-related luminosity, or even $0.1\,\%$ for several objects. The results imposes important constraints on any model of high energy radiation produced by Seyfert-type AGN.

There are several scenarios discussed in the literature in this context.
For example, as noted above, the star-forming activity taking place in the host galaxies of Seyfert objects should result in non-negligible production of $\gamma$ rays in the ISM. This inevitable emission component in Seyfert spectra is expected to be analogous to that observed by \lat in a few nearby star-forming galaxies \citep{SB}, and as such is expected to scale with the FIR and with the diffuse radio luminosities of the host galaxies. However, the flux level probed by the \lat in three years of all-sky survey does not allow for the detection of such diffuse emission for the majority of Seyferts, with a possible exception for the most nearby and actively star-forming targets. 
\citet{ten11} also mentioned the lack of detection of more distant Seyfert galaxies is likely a
\lat sensitivity issue based on their results of stacking analysis of 215 undetected Seyfert objects.
They derived the upper limit in the 1--100\,GeV energy range from the stacking analysis to be $\sim 3 \times 10^{41} {\rm erg}\,{\rm s}^{-1}$ assuming the median redshift of the 215 stacked objects ($z \sim 0.031$), but it is still approximately 3 and 18 times the $\gamma$-ray luminosities at 1--100\,GeV of
NGC~1068 and~NGC 4945, respectively.

In the previous section, we also commented on a possible contribution of radio jets to the $\gamma$-ray emission of Seyfert sources. Unlike the ISM-related $\gamma$-ray output, this jet-related emission component in Seyfert spectra at GeV photon energies is a subject of speculation \citep[see, e.g.,][]{len10,kat11}. Four AGN classified as NLS1s have been recently detected by \lat and their observed high-energy radiation was established to be due to the jet activity \citep{NLS,fos11b}. Those objects are however very different systems from the ones analyzed in this work, possessing flat-spectrum and high-brightness temperature radio cores, and therefore relativistic compact jets resembling blazar sources rather than sub-relativistic outflows observed in radio-quiet Seyferts \citep[e.g.,][]{fos09}. Since the targets studied in this paper should be considered as being representative for the whole population of such `classical' Seyferts, our analysis indicates that any jet-related $\gamma$-ray emission component in this type of AGN, even if present, is not as prominent as in radio galaxies or blazars.

Another possible emission site of $\gamma$-rays in Seyfert galaxies could be disk coronae, where the bulk of observed hard X-ray emission from Seyferts is produced. The first models for such emission involved non-thermal electron populations, and predicted power-law tails in Seyfert spectra extending to at least MeV photon energies \citep[e.g.][]{zdz85,sve87}. However, detections of spectral cut-offs in the hard X-ray continua around a few hundred keV photon energies in Seyfert galaxies favor emission models involving dominant thermal electron populations \citep[see the discussion in][]{pou98,zdz99}. Still, the available observational constraints do not exclude the presence of non-thermal power-law tails in Seyfert spectra in the MeV range, albeit with a much reduced normalization, constituting not more than $\sim 10\,\%$ of total energy radiated in the X-ray regime \citep{joh97,war01,lub10}.
Under this assumption, the observationally allowed luminosity ratio between the MeV ($0.1 -10$\,MeV) and the X-ray bands, $L_{\rm 0.1-10\,MeV}/L_{\rm X} < 0.1$, together with a simple scaling of the $0.1 - 10$\,GeV luminosity $L_{\gamma} = 10^{3\,(2-\Gamma)}\,L_{\rm 0.1-10\,MeV}$, formally implies an expected ratio $L_{\gamma}/L_{\rm X} < 0.003$ for $\Gamma = 2.5$. The current \lat sensitivity can hardly probe such levels of $\gamma$-ray emission at the moment.
In addition, even if the high energy emission were to originate in or near the accretion disk, the opacity of $\gamma$ rays to pair production via interaction with X-rays produced by the accretion disk might prevent those $\gamma$ rays from escaping.
Nevertheless, as pointed out by \citet{Ino08}, presence of such an emission component at the maximum allowed level would explain the observed extragalactic MeV background radiation in terms of a dominant contribution from Seyfert galaxies
while \citet{ten11} suggest the radio-quiet Seyfert galaxies are not a significant source of the extragalactic $\gamma$-ray background above 1\,GeV based on their analysis results of no $\gamma$-ray detection from the radio-quiet Seyfert galaxies at that energy range.

Finally, we discuss a possible mechanism of producing GeV photon in Seyfert galaxies by proton-proton interactions in the innermost parts of their accretion disks. Such a possibility was discussed previously in the context of Galactic black hole systems \citep{shap76,mah97,oka03}, and was applied recently to the case of active galaxies by \citet{nie09}. Although the current model predictions are still preliminary, this hadronic process was anticipated to result in a significant emission component in the $0.1-10$\,GeV range, possibly constituting $\gtrsim10\,\%$ of the disk/disk corona X-ray luminosity in the case of 
a particular (preferred) range of the accretion rate, typically corresponding 
to advection-dominated (``hot'') accretion flow, and of a maximally 
spinning black hole.  That is because for a Kerr black hole the innermost stable orbit of the accretion disk can be located much closer to the event horizon, and hence the number density of the matter within the innermost parts of the accretion disk as well as the proton temperature are increased, leading to enhanced proton-proton interactions above the threshold for the pion production. The \lat upper limits derived in this work for the sample of the hard X-ray--brightest Seyfert objects (mostly $L_{\gamma} / L_{\rm X} < 0.1$, and $<0.01$ in several particular cases) could be useful to constrain the model parameters and, ultimately, to determine the spin distribution for supermassive black holes hosted by Seyfert-type AGN. In this context, we find no prominent GeV emission component that could be related to hadronic interactions within accretion flows surrounding Kerr black holes for the whole analyzed sample, with the possible exceptions of ESO 323--G077 and NGC~6814.  

\begin{figure}
\epsscale{.80}
\plotone{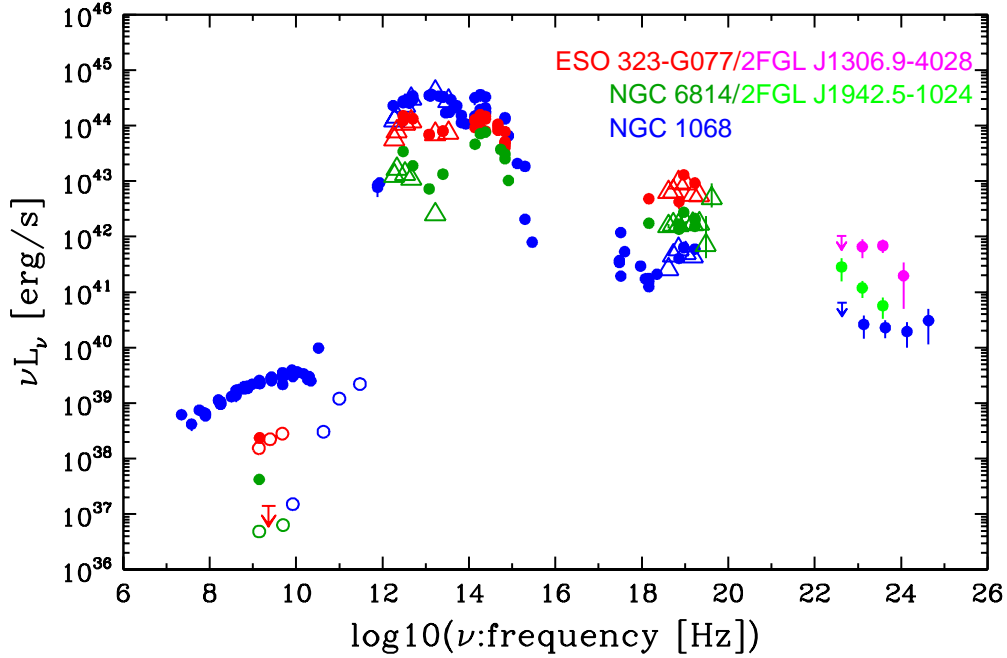}
\caption{Broad-band spectral energy distributions of ESO~323--G077 (red) and NGC~6814 (dark green). Both \lat spectra are derived when assuming the associations with 2FGL J1306.9--4028 (magenta) and 2FGL J1942.5--1024 (green), respectively.
For comparison, broad-band spectral energy distribution of the starburst galaxy NGC~1068 (blue) is also shown including its \lat data points from \citet{SB}. The data are taken from NED (total flux: filled circle), \akari and \bat (open triangle).
In the radio regime, core fluxes are also denoted as open circles for ESO~323-G077~\citep[from][]{cor02}, 
for NGC~6814~\citep[from][]{ulv84} and for NGC~1068 (from NED).
An upper limit for the radio compact core emission by  high-resolution ($<0''.05$) VLBI observations for ESO~323-G077 is also plotted~\citep[from][]{cor02}.
}
\label{fig:SED}
\end{figure}

As emphasized above, 
we cannot rule out the possibility that the associations of
2FGL~J1306.9--4028 with ESO 323--G077 and 2FGL J1942.5--1024 with NGC~6814 are due to chance spatial coincidences, 
but if the Seyfert objects are conclusively established as $\gamma$-ray emitters, 
then neither the star-forming nor the jet activity in these objects
can be considered as origins of the $\gamma$-ray
emission. Figure~\ref{fig:SED} shows the broad-band spectral energy
distribution of ESO~323--G077 (red data points), including the \lat
spectrum of 2FGL~J1306.9--4028 (magenta data points), as well as
NGC~6814 (dark green data points), and 2FGL J1942.5--1024 (green data points).
For comparison, in the figure we also plot the broad-band spectral energy distribution of the starburst galaxy NGC~1068 (blue data points). As shown in the figure and discussed in the previous section, the $\gamma$-ray--to--far-infrared luminosity ratio for ESO~323--G077/2FGL~J1306.9--4028 is much larger than for NGC~1068, which seems to exclude the possibility that the $\gamma$-ray emission from ESO~323--G077/2FGL~J1306.9--4028 is attributed to the star-forming activity within the host of the analyzed Seyfert. On one hand, no compact radio core is found in ESO~323--G077 even by high-resolution ($<0''.05$) VLBI observations, 
and the resultant upper limit for the radio core emission is 1.3\,mJy at 2.3~GHz \citep{cor02, cor03}. 
This finding challenges the jet hypothesis for the origin of $\gamma$ rays, if ESO~323--G077 is associated with 2FGL~J1306.9--4028.
Therefore, either the association of 2FGL~J1306.9--4028 with ESO
323--G077 is due to a chance positional coincidence, or ESO~323--G077
is an exceptional source for which the $\gamma$-ray radiative output
is dominated by an emission component not typically observed among
other Seyferts. The same reasoning may be applied in the case of NGC~6814.

\section{Conclusion}

In this paper, we report on a search for $\gamma$-ray emission from a sample of Seyfert galaxies selected via their hard X-ray fluxes, specifically for sources with $14-195$ keV fluxes above $2.5 \times 10^{-11}$ erg cm$^{-2}$ s$^{-1}$ as determined using the \bat 58-month catalog,
utilizing the three-year accumulation of \lat data.
We exclude `radio-loud' objects from the sample by selecting only those sources for which the parameter $R_{rX}$ --- the ratio of $\nu F_{\nu}$ radio flux at 1.4 GHz frequency to the hard X-ray flux in the $14-195$ keV band --- is less than $10^{-4}$. 
The selection criteria leave us with a well-defined sample of 120 `radio-quiet' Seyfert galaxies. 
The two nearby type-2 Seyferts which \textit{are} detected by the {\it Fermi}--LAT, NGC~1068 and NGC~4945, 
are not included in the analysis.
In a companion paper by \citet{SB}, 
we argue that the $\gamma$-ray emission of those two sources is more likely attributed to cosmic-ray interactions in the ISM of their host galaxies.

Generally, `radio-quiet' Seyfert galaxies selected by their hard X-ray
flux are \textit{not} detected in the $\gamma$-ray band covered by the
\lat. We report photon flux upper limits for all the sources included
in our sample: the typical limit is $\sim 4 \times 10^{-9}$ photons
cm$^{-2}$ s$^{-1}$ in the energy range above 100 MeV. We find two possible associations of $\gamma$-ray sources with
objects in our sample, ESO~323--G077 and NGC~6814, 
but caution that chance spatial coincidences with these objects cannot be ruled out.

FIR fluxes of the objects considered here, provided by the \akari satellite, indicate
the upper limits for the $L_{\gamma} / L_{\rm FIR}$ luminosity ratios in the range of $0.01 - 0.1$. 
At the same time, the LAT-detected starburst galaxies are characterized by $L_{\gamma} / L_{\rm FIR} \lesssim 0.001$. 
This suggests that detection of ISM emission in the GeV photon energy range from bulk of the hard X-ray selected Seyfert objects 
--- emission analogous to that observed in nearby star-forming galaxies --- 
would require increasing the sensitivity of the \lat survey by roughly an order of magnitude. 
Similarly, the derived upper limits for the $\gamma$-ray-to-radio luminosity ratio,  $L_{\gamma} / L_{\rm R} < 10^4$ on average, 
supports the conclusion by \citet{kat11} that the jet-related
$\gamma$-ray emission of Seyfert galaxies is generally expected to be
below the flux levels probed at present by \lat.

The resultant \lat upper limits yield the ratio of $\gamma$-ray
to X-ray luminosities $L_{\gamma} / L_{\rm X} < 0.1$, and even $<0.01$ in some cases.
In general,
coronae of accretion disks including
non-thermal electron populations can be considered as plausible sites of the $\gamma$-ray production.
Our analysis allows for the presence of such a broad-band power-law emission component extending from MeV to GeV range, but
constituting not more than 10\,\% of the thermal radiative output of the disks and disk coronae.

Finally, $\gamma$-ray photons may be produced in Seyfert galaxies by proton-proton interactions in the innermost parts of their accretion disks. 
Although the current model predictions are still preliminary, 
this hadronic process was anticipated to result in a significant emission component in the $0.1-10$\,GeV range, possibly constituting $\gtrsim10\,\%$ of the disk/disk corona X-ray luminosity in the case of a maximally spinning black hole. 
The upper limits derived in this paper indicate that no prominent GeV emission component that could be related to the hadronic interactions within accretion flows is found for the whole analyzed sample, with the possible exceptions of ESO~323--G077 and NGC~6814.

\acknowledgments
The \textit{Fermi}--LAT Collaboration acknowledges generous ongoing
support from a number of agencies and institutes that have supported
both the development and the operation of the LAT as well as
scientific data analysis. These include the National Aeronautics and
Space Administration and the Department of Energy in the United
States, the Commissariat \`a l'Energie Atomique and the Centre
National de la Recherche Scientifique / Institut National de Physique
Nucl\'eaire et de Physique des Particules in France, the Agenzia
Spaziale Italiana and the Istituto Nazionale di Fisica Nucleare in
Italy, the Ministry of Education, Culture, Sports, Science and
Technology (MEXT), High Energy Accelerator Research Organization (KEK)
and Japan Aerospace Exploration Agency (JAXA) in Japan, and the
K.~A.~Wallenberg Foundation, the Swedish Research Council and the
Swedish National Space Board in Sweden. Additional support for science
analysis during the operations phase is gratefully acknowledged from
the Istituto Nazionale di Astrofisica in Italy and the Centre National
d'\'Etudes Spatiales in France.

This research has made use of the NASA/IPAC Extragalactic Database (NED) 
which is operated by the Jet Propulsion Laboratory, California Institute of Technology, 
under contract with the National Aeronautics and Space Administration.

M.~H. is supported by the Research Fellowships of the Japan Society for the Promotion of Science for Young Scientists.
K.~B. is supported by a Stanford Graduate Fellowship.
We thank the anonymous referee for the valuable comments that helped to improve the paper.

{\it Facilities:} \facility{\lat}, \facility{\bat}, \facility{\akari}

\clearpage

\clearpage

\clearpage

\clearpage

\clearpage

\clearpage

\clearpage

\clearpage

\clearpage

\begin{deluxetable}{lrrlcrccccccl}
\tabletypesize{\tiny}
\rotate
\tablecaption{Basic information regarding Seyfert galaxies included in the analyzed sample \label{tab:samples}}
\tablewidth{0pt}
\tablehead{
\colhead{Name} & \colhead{R.A.} & \colhead{Dec }& \colhead{$z$} &  \colhead{$d_{\rm L}$} &
 \colhead{$F_{\rm 14-195\,keV}$} &  \colhead{$\log{L_{\rm X}}$} & \colhead{$\log{L_{\rm R}}$} & 
\colhead{ref.} & \colhead{$\log{R_{\rm rX}}$} & 
\colhead{$\log{L_{\rm FIR}}$} & \colhead{$\log{L_{\rm MIR}}$} & \colhead{type} \\    
 &  \colhead{[degree]}  &  \colhead{[degree]}  &    \colhead{} &  \colhead{[Mpc]} &
 \colhead{[$10^{-11}$\,cgs]}  &  \colhead{[erg s$^{-1}$]} & \colhead{[erg s$^{-1}$]} &  & & \colhead{[erg s$^{-1}$]} &  \colhead{[erg s$^{-1}$]} \\   
  \colhead{(1)} &   \colhead{(2)} &  \colhead{(3)} &  \colhead{(4)} &  \colhead{(5)} &  \colhead{(6)} &  \colhead{(7)} &  \colhead{(8)} &  \colhead{(9)} &  \colhead{(10)} 
  &  \colhead{(11)} &  \colhead{(12)}  &  \colhead{(13)} 
}
\startdata
Mrk 1501 & 2.6292  & 10.9749  & 0.08934 & 400 & 2.84  & 44.74  & 40.41  & 1 & -4.33  & \nodata & \nodata & Sy1.2 \\
NGC 235A & 10.7200  & -23.5410  & 0.02223 & 94.1 & 4.78  & 43.70  & 38.79  & 1 & -4.91  & 44.63  & \nodata & Sy1 \\
Mrk 348 & 12.1964  & 31.9570  & 0.01503 & 63.5 & 16.10  & 43.89  & 39.29  & 1 & -4.60  & 43.08  & \nodata & Sy2 \\
Mrk 1148 & 12.9783  & 17.4329  & 0.064 & 280.3 & 3.03  & 44.45  & 38.30  & 8 & -6.16  & \nodata & \nodata & Sy1 \\
Mrk 352 & 14.9720  & 31.8269  & 0.01486 & 62.7 & 2.89  & 43.13  & \nodata & \nodata & \nodata & \nodata & 44.04  & Sy1                   \\
Mrk 1152 & 18.4587  & -14.8456  & 0.05271 & 228.4 & 2.83  & 44.25  & 38.55  & 1 & -5.70  & \nodata & \nodata & Sy1.5 \\
Fairall 9 & 20.9408  & -58.8057  & 0.04702 & 205.3 & 5.05  & 44.41  & \nodata & \nodata & \nodata & \nodata & 44.59  & Sy1 \\
NGC 526A & 20.9766  & -35.0654  & 0.0191 & 80.9 & 5.95  & 43.67  & 38.15  & 1 & -5.51  & \nodata & 43.56  & Sy1.5 \\
ESO 297-018 & 24.6548  & -40.0114  & 0.0252 & 107.5 & 6.99  & 43.98  & 39.05  & 1 & -4.94  & 43.49  & \nodata & Sy2 \\
NGC 788 & 30.2769  & -6.8155  & 0.0136 & 56.3 & 8.31  & 43.50  & 37.25  & 2 & -6.25  & \nodata & \nodata & Sy2 \\
Mrk 1018 & 31.5666  & -0.2914  & 0.04244 & 181.5 & 3.24  & 44.11  & 38.36  & 1 & -5.75  & \nodata & \nodata & Sy1.5 \\
NGC 931 & 37.0603  & 31.3117  & 0.01665 & 50.4 & 6.08  & 43.27  & 37.75  & 1 & -5.52  & 43.39  & 43.55  & Sy1.5 \\
NGC 973 & 38.5838  & 32.5056  & 0.01619 & 60.5 & 2.85  & 43.10  & 38.12  & 1 & -4.97  & 43.39  & \nodata & Sy2 \\
NGC 985 & 38.6574  & -8.7876  & 0.043 & 184.7 & 3.11  & 44.10  & 38.94  & 1 & -5.17  & 44.24  & 44.34  & Sy1 \\
ESO 416-G002 & 38.8061  & -29.6047  & 0.0592 & 257.6 & 2.61  & 44.32  & 40.23  & 1 & -4.08  & \nodata & \nodata & Sy1.9 \\
ESO 198-024 & 39.5821  & -52.1923  & 0.0455 & 197.7 & 2.87  & 44.13  & 39.60  & 7 & -4.53  & \nodata & \nodata & Sy1 \\
2MASX J02485937+2630391 & 42.2472  & 26.5109  & 0.0579 & 252.4 & 3.45  & 44.42  & 39.37  & 1 & -5.05  & 44.55  & \nodata & Sy2 \\
MCG-02-08-014 & 43.0975  & -8.5104  & 0.01675 & 69.6 & 2.66  & 43.19  & 37.76  & 1 & -5.43  & \nodata & \nodata & Sy2 \\
NGC 1142 & 43.8008  & -0.1836  & 0.02885 & 121.5 & 9.52  & 44.23  & 39.58  & 1 & -4.65  & 44.62  & 44.20  & Sy2 \\
ESO 417-G006 & 44.0898  & -32.1856  & 0.01629 & 68.5 & 2.86  & 43.21  & 37.42  & 1 & -5.78  & \nodata & \nodata & Sy2 \\
NGC 1194 & 45.9546  & -1.1037  & 0.0136 & 56.2 & 3.72  & 43.15  & 37.12  & 1 & -6.03  & 42.82  & 43.33  & Sy1 \\
RX J0311.3-2046 & 47.8284  & -20.7717  & 0.066 & 293.8 & 2.77  & 44.46  & 38.81  & 1 & -5.64  & \nodata & \nodata & Sy1.5 \\
NGC 1365 & 53.4016  & -36.1404  & 0.00546 & 18.0 & 6.45  & 42.40  & 38.31  & 1 & -4.09  & 44.02  & 43.46  & Sy1.8 \\
ESO 548-G081 & 55.5155  & -21.2444  & 0.01448 & 60.4 & 4.20  & 43.26  & 37.29  & 1 & -5.97  & 43.15  & \nodata & Sy1 \\
ESO 549-G049 & 60.6070  & -18.0480  & 0.02629 & 111.1 & 2.57  & 43.58  & 38.71  & 1 & -4.87  & 44.17  & 43.90  & Sy2\\
UGC 03142 & 70.9450  & 28.9718  & 0.02166 & 91.7 & 4.95  & 43.70  & 38.51  & 1 & -5.19  & 43.72  & \nodata & Sy1 \\
2MASX J04440903+2813003 & 71.0376  & 28.2168  & 0.01127 & 47.7 & 6.03  & 43.22  & 38.04  & 1 & -5.18  & 43.11  & 42.91  & Sy2 \\
MCG-01-13-025 & 72.9229  & -3.8094  & 0.01589 & 66.7 & 3.17  & 43.23  & 37.76  & 1 & -5.47  & \nodata & \nodata & Sy1.2 \\
CGCG 420-015 & 73.3573  & 4.0616  & 0.02939 & 124.8 & 2.66  & 43.69  & 38.30  & 1 & -5.39  & 43.59  & 44.02  & Sy2 \\
2MASX J05054575-2351139 & 76.4405  & -23.8539  & 0.03504 & 150.3 & 6.19  & 44.22  & 38.47  & 1 & -5.75  & \nodata & \nodata & Sy2\\
CGCG 468-002NED01 & 77.0820  & 17.3630  & 0.0175 & 73.9 & 2.59  & 43.23  & 38.47  & 1 & -4.76  & 44.38  & \nodata & Sy2 \\
IRAS 05078+1626 & 77.6896  & 16.4989  & 0.01788 & 75.5 & 8.88  & 43.78  & 37.76  & 1 & -6.03  & 43.34  & 43.59  & Sy1.5 \\
2MASX J05151978+1854515 & 78.8324  & 18.9143  & \nodata & \nodata & 3.31  & \nodata & \nodata & 1 & -5.28  & \nodata & \nodata & Galaxy \\
Ark 120 & 79.0476  & -0.1498  & 0.0323 & 139.7 & 6.63  & 44.19  & 38.60  & 1 & -5.59  & \nodata & 44.29  & Sy1 \\
ESO 362-18 & 79.8993  & -32.6578  & 0.01245 & 53.0 & 5.11  & 43.23  & 37.82  & 1 & -5.41  & 43.16  & 43.28  & Sy1.5 \\
2MASX J05442257+5907361 & 86.0941  & 59.1267  & 0.06597 & 292.9 & 2.65  & 44.43  & 39.23  & 1 & -5.20  & \nodata & \nodata & Sy1.9 \\
NGC 2110 & 88.0474  & -7.4562  & 0.00779 & 29.0 & 29.74  & 43.48  & 38.62  & 1 & -4.85  & 43.19  & 43.00  & Sy2 \\
MCG+08-11-011 & 88.7234  & 46.4393  & 0.02048 & 88.3 & 13.05  & 44.09  & 39.50  & 1 & -4.58  & 43.87  & 44.02  & Sy1.5 \\
2MASX J05580206-3820043 & 89.5083  & -38.3346  & 0.03387 & 146.7 & 2.90  & 43.87  & 39.09  & 1 & -4.78  & \nodata & 44.48  & Sy1 \\
ESO 005-G004 & 91.4235  & -86.6319  & 0.00623 & 22.0 & 3.59  & 42.32  & 37.85  & 7 & -4.47  & 43.21  & 43.02  & Sy2 \\
ESO 121-IG028 & 95.9400  & -60.9790  & 0.0403 & 177.8 & 2.69  & 44.01  & \nodata & \nodata & \nodata & \nodata & \nodata & Sy2 \\
ESO 426-G002 & 95.9434  & -32.2166  & 0.02243 & 97.1 & 2.66  & 43.48  & \nodata & \nodata & \nodata & \nodata & \nodata & Sy2 \\
ESO 490-IG026 & 100.0487  & -25.8954  & 0.0248 & 107.7 & 3.78  & 43.72  & 38.87  & 1 & -4.85  & 43.81  & \nodata & Sy1.2 \\
2MASX J06411806+3249313 & 100.3252  & 32.8254  & 0.047 & 205.2 & 3.67  & 44.27  & 38.66  & 1 & -5.61  & \nodata & \nodata & Sy2 \\
Mrk 6 & 103.0511  & 74.4271  & 0.01881 & 83.0 & 6.20  & 43.71  & 39.49  & 1 & -4.22  & 43.40  & 43.69  & Sy1.5 \\
Mrk 79 & 115.6367  & 49.8097  & 0.02219 & 97.3 & 4.64  & 43.72  & 38.51  & 1 & -5.21  & 43.71  & 44.02  & Sy1.2 \\
2MASX J07595347+2323241 & 119.9728  & 23.3901  & 0.02918 & 127.7 & 3.19  & 43.79  & 38.78  & 1 & -5.02  & 44.26  & 43.85  & Sy2 \\
IC 0486 & 120.0874  & 26.6135  & 0.02688 & 112.0 & 3.58  & 43.73  & 38.33  & 1 & -5.40  & 43.78  & \nodata & Sy1 \\
Mrk 1210 & 121.0244  & 5.1138  & 0.0135 & 60.3 & 5.31  & 43.36  & 38.84  & 1 & -4.52  & 43.24  & 43.59  & Sy2 \\
Fairall 272 & 125.7546  & -4.9349  & 0.02182 & 96.7 & 4.76  & 43.73  & 38.76  & 1 & -4.96  & 43.47  & \nodata & Sy2 \\
Mrk 704 & 139.6084  & 16.3053  & 0.02923 & 130.0 & 3.28  & 43.82  & 38.23  & 1 & -5.59  & \nodata & 44.24  & Sy1.5 \\
MCG-01-24-012 & 140.1927  & -8.0561  & 0.01964 & 89.0 & 4.12  & 43.59  & 38.54  & 1 & -5.05  & \nodata & \nodata & Sy2 \\
MCG+04-22-042 & 140.9292  & 22.9090  & 0.03235 & 143.6 & 4.18  & 44.01  & 38.55  & 1 & -5.47  & \nodata & 43.82  & Sy1.2 \\
Mrk 110 & 141.3036  & 52.2863  & 0.03529 & 156.0 & 5.63  & 44.21  & 38.59  & 1 & -5.62  & \nodata & \nodata & Sy1 \\
MCG-05-23-016 & 146.9173  & -30.9489  & 0.00849 & 36.8 & 19.80  & 43.51  & 37.51  & 1 & -6.00  & 42.84  & 43.31  & Sy2 \\
NGC 3081 & 149.8731  & -22.8263  & 0.00796 & 28.6 & 8.44  & 42.92  & 36.87  & 1 & -6.05  & 42.89  & 42.74  & Sy2 \\
ESO 263-G013 & 152.4509  & -42.8112  & 0.03329 & 150.9 & 3.34  & 43.96  & 38.79  & 7 & -5.17  & \nodata & \nodata & Sy2 \\
NGC 3227 & 155.8774  & 19.8651  & 0.00386 & 26.4 & 11.28  & 42.97  & 38.06  & 1 & -4.92  & 43.47  & 43.09  & Sy1.5 \\
NGC 3281 & 157.9670  & -34.8537  & 0.01067 & 46.4 & 8.71  & 43.35  & 38.46  & 1 & -4.89  & 43.71  & 43.55  & Sy2 \\
2MASS J10315431-1416514 & 157.9763  & -14.2809  & 0.086 & 387.3 & 3.42  & 44.79  & 39.54  & 1 & -5.24  & \nodata & 44.73  & Sy1 \\
NGC 3393 & 162.0977  & -25.1621  & 0.01251 & 57.4 & 2.55  & 43.00  & 38.65  & 1 & -4.36  & 43.44  & \nodata & Sy2 \\
Mrk 417 & 162.3789  & 22.9644  & 0.03276 & 147.4 & 3.36  & 43.94  & \nodata & \nodata & \nodata & \nodata & \nodata & Sy2 \\
NGC 3516 & 166.6979  & 72.5686  & 0.00884 & 38.0 & 12.31  & 43.33  & 37.88  & 1 & -5.45  & 42.88  & 43.18  & Sy1.5 \\
NGC 3783 & 174.7572  & -37.7386  & 0.00973 & 25.1 & 18.77  & 43.15  & 37.66  & 1 & -5.49  & 42.83  & 43.10  & Sy1 \\
UGC 06728 & 176.3168  & 79.6815  & 0.00652 & 32.9 & 2.68  & 42.54  & \nodata & \nodata & \nodata & \nodata & \nodata & Sy1.2 \\
2MASX J11454045-1827149 & 176.4186  & -18.4543  & 0.03295 & 150.7 & 4.95  & 44.13  & 38.55  & 1 & -5.57  & \nodata & \nodata & Sy1 \\
NGC 4051 & 180.7901  & 44.5313  & 0.00233 & 17.1 & 3.76  & 42.12  & 37.66  & 1 & -4.45  & 42.73  & 42.61  & Sy1.5 \\
ARK 347 & 181.1237  & 20.3162  & 0.02244 & 104.1 & 2.92  & 43.58  & 37.90  & 1 & -5.68  & \nodata & \nodata & Sy2 \\
NGC 4138 & 182.3741  & 43.6853  & 0.00296 & 13.8 & 3.07  & 41.84  & 36.78  & 1 & -5.06  & 42.21  & \nodata & Sy1.9 \\
NGC 4151 & 182.6357  & 39.4057  & 0.00332 & 11.2 & 53.31  & 42.90  & 37.88  & 1 & -5.03  & 42.36  & 42.71  & Sy1.5 \\
NGC 4235 & 184.2912  & 7.1916  & 0.00804 & 31.5 & 3.14  & 42.57  & 37.31  & 1 & -5.27  & 42.20  & \nodata & Sy1 \\
NGC 4388 & 186.4448  & 12.6621  & 0.00842 & 16.8 & 27.58  & 42.97  & 37.75  & 1 & -5.22  & 43.07  & 42.71  & Sy2 \\
NGC 4395 & 186.4538  & 33.5468  & 0.00106 & 4.74 & 2.61  & 40.85  & 34.65  & 2 & -6.19  & 41.16  & \nodata & Sy1.9 \\
NGC 4507 & 188.9026  & -39.9093  & 0.0118 & 62.4 & 19.04  & 43.95  & 38.63  & 1 & -5.31  & 43.83  & 43.90  & Sy2 \\
ESO 506-G027 & 189.7275  & -27.3078  & 0.02502 & 119.0 & 9.26  & 44.20  & 39.24  & 1 & -4.96  & 43.54  & 43.79  & Sy2 \\
LEDA 170194 & 189.7762  & -16.1797  & 0.03667 & 167.7 & 4.37  & 44.17  & 39.26  & 1 & -4.90  & \nodata & \nodata & Sy2 \\
NGC 4593 & 189.9143  & -5.3443  & 0.009 & 37.3 & 8.87  & 43.17  & 37.01  & 1 & -6.16  & \nodata  & \nodata & Sy1 \\
NGC 4686 & 191.6661  & 54.5342  & 0.01674 & 77.8 & 2.79  & 43.31  & 37.64  & 1 & -5.67  & \nodata & \nodata & Galaxy \\
SBS 1301+540 & 195.9978  & 53.7917  & 0.02988 & 134.8 & 3.46  & 43.88  & 37.91  & 1 & -5.96  & \nodata & \nodata & Sy1 \\
NGC 4939 & 196.0600  & -10.3396  & 0.01037 & 34.7 & 2.54  & 42.56  & 37.85  & 1 & -4.72  & 43.09  & \nodata & Sy2 \\
ESO 323-G077 & 196.6089  & -40.4146  & 0.01501 & 62.4 & 4.25  & 43.30  & 38.36  & 4 & -4.94  & 44.03  & 43.86  & Sy1.2 \\
NGC 4992 & 197.2733  & 11.6341  & 0.02514 & 117.0 & 5.56  & 43.96  & 37.66  & 2 & -6.30  & 43.28  & 43.52  & Sy2 \\
MCG-06-30-015 & 203.9741  & -34.2956  & 0.00775 & 25.5 & 6.36  & 42.69  & 36.27  & 5 & -6.43  & 42.43  & 42.86  & Sy1.2 \\
NGC 5252 & 204.5665  & 4.5426  & 0.02297 & 108.4 & 11.11  & 44.19  & 38.50  & 1 & -5.69  & 43.29  & \nodata & Sy1.9 \\
IC 4329A & 207.3303  & -30.3094  & 0.01605 & 83.0 & 28.96  & 44.38  & 38.88  & 1 & -5.50  & 43.69  & 44.33  & Sy1.2 \\
Mrk 279 & 208.2644  & 69.3082  & 0.03045 & 136.0 & 4.42  & 43.99  & 38.85  & 1 & -5.14  & \nodata & 44.01  & Sy1.5 \\
NGC 5506 & 213.3119  & -3.2075  & 0.00618 & 21.7 & 24.28  & 43.14  & 38.43  & 1 & -4.71  & 43.20  & 43.19  & Sy1.9 \\
NGC 5548 & 214.4981  & 25.1368  & 0.01717 & 82.2 & 7.36  & 43.77  & 38.50  & 1 & -5.27  & 43.46  & 43.63  & Sy1.5 \\
ESO 511-G030 & 214.8434  & -26.6447  & 0.02239 & 108.9 & 4.41  & 43.80  & 38.39  & 1 & -5.40  & 43.60  & \nodata & Sy1 \\
Mrk 817 & 219.0920  & 58.7943  & 0.03145 & 141.5 & 2.74  & 43.82  & 38.57  & 1 & -5.25  & 44.10  & 44.18  & Sy1.5 \\
NGC 5728 & 220.5997  & -17.2532  & 0.0093 & 24.8 & 9.09  & 42.83  & 37.86  & 1 & -4.97  & 43.45  & 42.64  & Sy2 \\
IC 4518A & 224.4216  & -43.1321  & 0.01626 & 82.0 & 2.79  & 43.35  & 39.20  & 4 & -4.15  & 44.28  & 43.81  & Sy2 \\
Mrk 841 & 226.0050	& 10.4378 & 	0.03642 &	162.1 & 3.61 & 44.08 & \nodata & \nodata & \nodata & \nodata & 44.16 & Sy1 \\                  
2MASX J15115979-2119015 & 227.9992  & -21.3171  & 0.04461 & 203.2 & 3.24  & 44.20  & 39.51  & 1 & -4.70  & 44.42  & 44.17  & Sy1/NL \\
2MASX J15144217-8123377 & 228.6751  & -81.3939  & 0.06837 & 306.9 & 3.09  & 44.54  & 39.09  & 7 & -5.45  & \nodata & \nodata & Sy1.2 \\
MCG-01-40-001 & 233.3363  & -8.7005  & 0.02271 & 107.5 & 3.27  & 43.65  & 39.61  & 1 & -4.04  & 44.03  & \nodata & Sy2 \\
NGC 5995 & 237.1040  & -13.7578  & 0.02519 & 118.1 & 4.16  & 43.84  & 38.84  & 1 & -5.00  & 44.41  & 44.26  & Sy2 \\
Mrk 1498 & 247.0169  & 51.7754  & 0.0547 & 245.8 & 4.24  & 44.49  & 39.60  & 1 & -4.88  & \nodata & 44.23  & Sy1.9 \\
NGC 6240 & 253.2454  & 2.4009  & 0.02448 & 113.5 & 6.70  & 44.01  & 39.96  & 1 & -4.05  & 45.08  & 44.25  & Sy2 \\
NGC 6300 & 259.2478  & -62.8206  & 0.0037 & 13.1 & 9.70  & 42.30  & 37.31  & 7 & -4.99  & 43.01  & 42.28  & Sy2 \\
2MASX J18074992+1120494 & 271.9580  & 11.3470  & \nodata & \nodata & 2.84  & \nodata & \nodata & \nodata & \nodata & \nodata & \nodata & Galaxy \\
ESO 103-035 & 279.5848  & -65.4276  & 0.01329 & 60.5 & 11.31  & 43.69  & 38.17  & 7 & -5.52  & 43.25  & 43.64  & Sy2 \\
Fairall 51 & 281.2249  & -62.3648  & 0.01417 & 64.1 & 4.26  & 43.32  & 37.88  & 7 & -5.44  & 43.45  & 43.69  & Sy1 \\
ESO 141-G055 & 290.3090  & -58.6703  & 0.036 & 165.4 & 5.34  & 44.24  & 38.44  & 7 & -5.81  & \nodata & 44.21  & Sy1 \\
NGC 6814 & 295.6694  & -10.3235  & 0.00521 & 22.0 & 7.53  & 42.64  & 37.60  & 1 & -5.03  & 43.13  & \nodata & Sy1.5 \\
NGC 6860 & 302.1954  & -61.1002  & 0.01488 & 67.5 & 5.28  & 43.46  & 38.00  & 7 & -5.46  & 43.40  & 43.44  & Sy1 \\
Mrk 509 & 311.0406  & -10.7235  & 0.0344 & 151.6 & 9.42  & 44.41  & 38.85  & 1 & -5.56  & \nodata & 44.36  & Sy1.2 \\
6dF J2132022-334254 & 323.0092  & -33.7150  & 0.02929 & 131.4 & 4.45  & 43.96  & 37.96  & 1 & -6.00  & \nodata & \nodata & Sy1 \\
1RXS J213623.1-622400 & 324.0963  & -62.4002  & 0.0588 & 260.0 & 2.92  & 44.37  & 38.70  & 7 & -5.68  & \nodata & \nodata & Sy1 \\
Mrk 520 & 330.1724  & 10.5524  & 0.02661 & 115.5 & 3.18  & 43.70  & 39.11  & 1 & -4.60  & 44.43  & 43.90  & Sy1.9 \\
NGC 7172 & 330.5080  & -31.8698  & 0.00868 & 31.9 & 17.36  & 43.32  & 37.80  & 1 & -5.53  & 43.52  & 43.11  & Sy2 \\
NGC 7213 & 332.3177  & -47.1667  & 0.00584 & 14.5 & 4.43  & 42.05  & 37.47  & 7 & -4.58  & 42.39  & 42.48  & Sy1.5 \\
NGC 7314 & 338.9426  & -26.0503  & 0.00476 & 15.9 & 5.12  & 42.19  & 37.12  & 1 & -5.07  & 42.66  & \nodata & Sy1.9 \\
NGC 7319 & 339.0148  & 33.9757  & 0.02251 & 97.3 & 3.93  & 43.65  & 38.92  & 1 & -4.73  & 43.34  & 43.53  & Sy2 \\
Mrk 915 & 339.1938  & -12.5452  & 0.02411 & 104.0 & 3.22  & 43.62  & 39.10  & 1 & -4.52  & \nodata  & \nodata & Sy1 \\
MR 2251-178 & 343.5242  & -17.5819  & 0.06398 & 282.3 & 10.03  & 44.98  & 39.33  & 1 & -5.65  & \nodata & \nodata & Sy1 \\
NGC 7469 & 345.8151  & 8.8740  & 0.01632 & 69.9 & 6.87  & 43.60  & 39.17  & 1 & -4.44  & 44.73  & 44.18  & Sy1.2 \\
Mrk 926 & 346.1811  & -8.6857  & 0.04686 & 203.8 & 11.25  & 44.75  & 39.35  & 1 & -5.40  & 44.03  & 44.00  & Sy1.5 \\
NGC 7582 & 349.5979  & -42.3706  & 0.00525 & 18.7 & 8.10  & 42.53  & 38.19  & 4 & -4.34  & 43.93  & 43.28  & Sy2 \\
NGC 7603 & 349.7359  & 0.2440  & 0.02952 & 126.5 & 4.85  & 43.97  & 38.81  & 1 & -5.16  & 43.93  & 44.28  & Sy1.5 \\
\enddata
\tablecomments{(1) source name from the \bat catalog; 
(2) J2000;
(3) J2000; 
(4) redshift;
(5) luminosity distance;
(6) $14 -195$\,keV energy flux from the \bat 58-month catalog; 
(7) $14 -195$\,keV luminosity;
(8) $1.4$\,GHz radio luminosity; 
(9) references to radio data; 
(10) hard X-ray radio loudness parameter;
(11) FIR luminosity at 90\,\micron\ from  the\akari$\!\!$--FIS data; 
(12) MIR luminosity at 9\,\micron\ from the \akari$\!\!$--IRC data;  
(13) source type as given in the 58-month \bat catalog \citep{Bau10}.}
\tablerefs{
1. \citet{NVSS} (NVSS); 
2. \citet{FIRST} (FIRST, Version 03Apr11); 
3. \citet{PKS} (PKSCAT90) ; 
4. \citet{con96}; 
5. \citet{ulv84}; 
6. \citet{whi92};
7. \citet{SUMSS} (SUMSS V2.1: 0.843\,GHz);
8. \citet{Mil93} (4.86\,GHz)
}
\end{deluxetable}

\clearpage

\begin{deluxetable}{lrrrcccl}
\tabletypesize{\tiny}
\rotate
\tablecaption{Results of the \lat data analysis for the selected sample of Seyfert galaxies \label{tab:result}}
\tablewidth{0pt}
\tablehead{
\colhead{Name} &   \colhead{R.A.} & \colhead{Dec }& \colhead{$TS$} & 
\colhead{UL: $\mathcal{F}(\rm >0.1\,GeV)$} & \colhead{UL2: $\mathcal{F}(\rm >0.1\,GeV)$} & \colhead{UL: $\log{L_{\gamma}}$} & \colhead{UL: $L_{\gamma}/L_{\rm X}$}  \\    
  &  \colhead{[degree]}  &  \colhead{[degree]}  & 
 & \colhead{[$10^{-9}$\,ph cm$^{-2}$ s$^{-1}$]}  &  \colhead{[$10^{-9}$\,ph cm$^{-2}$ s$^{-1}$]}  &  
 \colhead{[erg s$^{-1}$]}  &    \\   
  \colhead{(1)} &   \colhead{(2)} &  \colhead{(3)} &  \colhead{(4)} &  \colhead{(5)} &  \colhead{(6)} &  \colhead{(7)} & \colhead{(8)}
}
\startdata
Mrk 1501 & 2.6292  & $10.9749 $ & 0.0  & 4.3  & 2.6  & 43.6  & 0.065 \\
NGC 235A & 10.7200  & $-23.5410 $ & 9.1  & 7.7  & 5.2  & 42.6  & 0.070 \\
Mrk 348 & 12.1964  & $31.9570 $ & 0.0  & 2.0  & 1.3  & 41.6  & 0.0055 \\
Mrk 1148 & 12.9783  & $17.4329 $ & 0.0  & 3.0  & 1.6  & 43.1  & 0.043 \\
Mrk 352 & 14.9720  & $31.8269 $ & 0.0  & 3.2  & 2.1  & 41.8  & 0.048 \\
Mrk 1152 & 18.4587  & $-14.8456 $ & 1.3  & 3.3  & 1.9  & 43.0  & 0.051 \\
Fairall 9 & 20.9408  & $-58.8057 $ & 0.0  & 2.3  & 1.5  & 42.7  & 0.020 \\
NGC 526A & 20.9766  & $-35.0654 $ & 0.0  & 1.4  & 0.82  & 41.7  & 0.010 \\
ESO 297-018 & 24.6548  & $-40.0114 $ & 2.5  & 4.8  & 3.1  & 42.5  & 0.030 \\
NGC 788 & 30.2769  & $-6.8155 $ & 2.7  & 4.6  & 2.9  & 41.9  & 0.024 \\
Mrk 1018 & 31.5666  & $-0.2914 $ & 0.7  & 3.9  & 2.1  & 42.8  & 0.052 \\
NGC 931 & 37.0603  & $31.3117 $ & 6.4  & 9.4  & 5.5  & 42.1  & 0.067 \\
NGC 973 & 38.5838  & $32.5056 $ & 0.0  & 3.5  & 1.9  & 41.8  & 0.054 \\
NGC 985 & 38.6574  & $-8.7876 $ & 0.0  & 2.6  & 1.8  & 42.7  & 0.037 \\
ESO 416-G002 & 38.8061  & $-29.6047 $ & 8.5  & 4.9  & 3.6  & 43.2  & 0.082 \\
ESO 198-024 & 39.5821  & $-52.1923 $ & 0.0  & 3.2  & 2.0  & 42.8  & 0.049 \\
2MASX J02485937+2630391 & 42.2472  & $26.5109 $ & 0.0  & 3.5  & 2.1  & 43.1  & 0.043 \\
MCG-02-08-014 & 43.0975  & $-8.5104 $ & 0.9  & 2.5  & 2.0  & 41.8  & 0.041 \\
NGC 1142 & 43.8008  & $-0.1836 $ & 0.0  & 1.8  & 1.1  & 42.1  & 0.0082 \\
ESO 417-G006 & 44.0898  & $-32.1856 $ & 0.0  & 1.8  & 1.2  & 41.6  & 0.027 \\
NGC 1194 & 45.9546  & $-1.1037 $ & 6.2  & 3.9  & 3.1  & 41.8  & 0.045 \\
RX J0311.3-2046 & 47.8284  & $-20.7717 $ & 0.0  & 3.3  & 2.0  & 43.2  & 0.052 \\
NGC 1365 & 53.4016  & $-36.1404 $ & 1.1  & 4.6  & 2.6  & 40.9  & 0.031 \\
ESO 548-G081 & 55.5155  & $-21.2444 $ & 0.0  & 5.2  & 2.6  & 42.0  & 0.054 \\
ESO 549-G049 & 60.6070  & $-18.0480 $ & 1.1  & 4.7  & 2.5  & 42.5  & 0.079 \\
UGC 03142 & 70.9450  & $28.9718 $ & 1.6  & 9.0  & 6.3  & 42.6  & 0.079 \\
2MASX J04440903+2813003 & 71.0376  & $28.2168 $ & 7.4  & 11.9  & 8.0  & 42.1  & 0.085 \\
MCG-01-13-025 & 72.9229  & $-3.8094 $ & 3.2  & 3.1  & 2.3  & 41.9  & 0.043 \\
CGCG 420-015 & 73.3573  & $4.0616 $ & 0.0  & 3.6  & 2.1  & 42.5  & 0.059 \\
2MASX J05054575-2351139 & 76.4405  & $-23.8539 $ & 4.2  & 7.8  & 4.2  & 43.0  & 0.055 \\
CGCG 468-002NED01 & 77.0820  & $17.3630 $ & 0.0  & 6.1  & 3.0  & 42.2  & 0.10 \\
IRAS 05078+1626 & 77.6896  & $16.4989 $ & 0.0  & 3.1  & 1.7  & 42.0  & 0.015 \\
2MASX J05151978+1854515 & 78.8324  & $18.9143 $ & 0.0  & 5.4  & 3.1  & \nodata & 0.071 \\
Ark 120 & 79.0476  & $-0.1498 $ & 0.0  & 2.4  & 1.6  & 42.4  & 0.016 \\
ESO 362-18 & 79.8993  & $-32.6578 $ & 7.1  & 8.0  & 4.6  & 42.1  & 0.068 \\
2MASX J05442257+5907361 & 86.0941  & $59.1267 $ & 0.0  & 2.9  & 1.9  & 43.1  & 0.048 \\
NGC 2110 & 88.0474  & $-7.4562 $ & 0.1  & 5.2  & 2.2  & 41.4  & 0.0076 \\
MCG+08-11-011 & 88.7234  & $46.4393 $ & 0.0  & 3.3  & 1.9  & 42.1  & 0.011 \\
2MASX J05580206-3820043 & 89.5083  & $-38.3346 $ & 0.0  & 7.1  & 3.5  & 42.9  & 0.11 \\
ESO 005-G004 & 91.4235  & $-86.6319 $ & 0.0  & 2.8  & 1.6  & 40.8  & 0.033 \\
ESO 121-IG028 & 95.9400  & $-60.9790 $ & 0.0  & 2.2  & 1.2  & 42.6  & 0.036 \\
ESO 426-G002 & 95.9434  & $-32.2166 $ & 0.7  & 5.9  & 3.4  & 42.5  & 0.095 \\
ESO 490-IG026 & 100.0487  & $-25.8954 $ & 0.1  & 3.4  & 2.6  & 42.3  & 0.039 \\
2MASX J06411806+3249313 & 100.3252  & $32.8254 $ & 0.5  & 4.7  & 3.3  & 43.0  & 0.055 \\
Mrk 6 & 103.0511  & $74.4271 $ & 0.0  & 3.7  & 2.3  & 42.1  & 0.026 \\
Mrk 79 & 115.6367  & $49.8097 $ & 2.5  & 6.1  & 3.2  & 42.5  & 0.057 \\
2MASX J07595347+2323241 & 119.9728  & $23.3901 $ & 0.3  & 3.9  & 2.2  & 42.5  & 0.052 \\
IC 0486 & 120.0874  & $26.6135 $ & 10.6  & 9.3  & 6.3  & 42.8  & 0.11 \\
Mrk 1210 & 121.0244  & $5.1138 $ & 4.4  & 7.8  & 4.5  & 42.2  & 0.063 \\
Fairall 272 & 125.7546  & $-4.9349 $ & 0.1  & 3.3  & 2.1  & 42.2  & 0.030 \\
Mrk 704 & 139.6084  & $16.3053 $ & 0.0  & 3.5  & 2.0  & 42.5  & 0.046 \\
MCG-01-24-012 & 140.1927  & $-8.0561 $ & 0.0  & 2.3  & 1.5  & 42.0  & 0.024 \\
MCG+04-22-042 & 140.9292  & $22.9090 $ & 0.0  & 2.4  & 1.6  & 42.4  & 0.025 \\
Mrk 110 & 141.3036  & $52.2863 $ & 0.0  & 3.0  & 1.9  & 42.6  & 0.023 \\
MCG-05-23-016 & 146.9173  & $-30.9489 $ & 0.0  & 2.4  & 1.5  & 41.2  & 0.0053 \\
NGC 3081 & 149.8731  & $-22.8263 $ & 0.0  & 4.2  & 2.5  & 41.2  & 0.021 \\
ESO 263-G013 & 152.4509  & $-42.8112 $ & 0.0  & 5.6  & 3.6  & 42.8  & 0.073 \\
NGC 3227 & 155.8774  & $19.8651 $ & 3.1  & 5.2  & 2.6  & 41.3  & 0.020 \\
NGC 3281 & 157.9670  & $-34.8537 $ & 3.3  & 8.0  & 5.1  & 41.9  & 0.040 \\
2MASS J10315431-1416514 & 157.9763  & $-14.2809 $ & 8.9  & 7.5  & 5.2  & 43.8  & 0.094 \\
NGC 3393 & 162.0977  & $-25.1621 $ & 8.2  & 5.8  & 4.3  & 42.0  & 0.098 \\
Mrk 417 & 162.3789  & $22.9644 $ & 1.2  & 6.5  & 3.7  & 42.9  & 0.084 \\
NGC 3516 & 166.6979  & $72.5686 $ & 4.3  & 6.8  & 3.4  & 41.7  & 0.024 \\
NGC 3783 & 174.7572  & $-37.7386 $ & 3.3  & 7.5  & 4.7  & 41.4  & 0.017 \\
UGC 06728 & 176.3168  & $79.6815 $ & 0.0  & 2.3  & 1.4  & 41.1  & 0.037 \\
2MASX J11454045-1827149 & 176.4186  & $-18.4543 $ & 0.6  & 5.1  & 2.8  & 42.8  & 0.045 \\
NGC 4051 & 180.7901  & $44.5313 $ & 0.0  & 2.7  & 1.7  & 40.6  & 0.031 \\
Ark 347 & 181.1237  & $20.3162 $ & 5.0  & 6.7  & 3.1  & 42.6  & 0.10 \\
NGC 4138 & 182.3741  & $43.6853 $ & 1.5  & 4.3  & 2.7  & 40.6  & 0.061 \\
NGC 4151 & 182.6357  & $39.4057 $ & 0.0  & 3.1  & 2.1  & 40.3  & 0.0025 \\
NGC 4235 & 184.2912  & $7.1916 $ & 0.7  & 5.3  & 3.2  & 41.4  & 0.073 \\
NGC 4388 & 186.4448  & $12.6621 $ & 0.0  & 3.5  & 2.1  & 40.7  & 0.0055 \\
NGC 4395 & 186.4538  & $33.5468 $ & 0.0  & 3.3  & 2.1  & 39.6  & 0.055 \\
NGC 4507 & 188.9026  & $-39.9093 $ & 1.1  & 6.8  & 4.1  & 42.1  & 0.016 \\
ESO 506-G027 & 189.7275  & $-27.3078 $ & 0.0  & 3.0  & 1.8  & 42.3  & 0.014 \\
LEDA 170194 & 189.7762  & $-16.1797 $ & 0.3  & 5.3  & 3.1  & 42.9  & 0.052 \\
NGC 4593 & 189.9143  & $-5.3443 $ & 5.1  & 7.6  & 3.6  & 41.7  & 0.037 \\
NGC 4686 & 191.6661  & $54.5342 $ & 0.0  & 2.0  & 1.3  & 41.8  & 0.032 \\
SBS 1301+540 & 195.9978  & $53.7917 $ & 0.3  & 3.9  & 2.5  & 42.6  & 0.049 \\
NGC 4939 & 196.0600  & $-10.3396 $ & 0.1  & 4.0  & 1.9  & 41.4  & 0.067 \\
ESO 323-G077 & 196.6089  & $-40.4146 $ & 0.0\tablenotemark{a}  & 6.5  & 5.3  & 42.1  & 0.066 \\
  &  & &26.7 & \multicolumn{2}{c}{[Flux] $8.2\pm2.7$\tablenotemark{b}, [Index] $\Gamma=2.21\pm0.14$\tablenotemark{b}} & 42.3 & 0.11\\
NGC 4992 & 197.2733  & $11.6341 $ & 0.0  & 3.9  & 2.7  & 42.4  & 0.030 \\
MCG-06-30-015 & 203.9741  & $-34.2956 $ & 0.0  & 4.2  & 2.4  & 41.2  & 0.029 \\
NGC 5252 & 204.5665  & $4.5426 $ & 0.0  & 2.7  & 1.4  & 42.2  & 0.010 \\
IC 4329A & 207.3303  & $-30.3094 $ & 2.5  & 7.6  & 4.6  & 42.4  & 0.011 \\
Mrk 279 & 208.2644  & $69.3082 $ & 3.6  & 5.1  & 3.0  & 42.7  & 0.050 \\
NGC 5506 & 213.3119  & $-3.2075 $ & 0.0  & 4.0  & 2.4  & 41.0  & 0.0071 \\
NGC 5548 & 214.4981  & $25.1368 $ & 0.0  & 3.7  & 2.4  & 42.1  & 0.022 \\
ESO 511-G030 & 214.8434  & $-26.6447 $ & 0.0  & 2.6  & 1.6  & 42.2  & 0.025 \\
Mrk 817 & 219.0920  & $58.7943 $ & 0.0  & 2.4  & 1.6  & 42.4  & 0.038 \\
NGC 5728 & 220.5997  & $-17.2532 $ & 2.2  & 7.5  & 3.9  & 41.4  & 0.035 \\
IC 4518A & 224.4216  & $-43.1321 $ & 0.6  & 7.3  & 3.5  & 42.4  & 0.11 \\
Mrk 841 & 226.0050  & $10.4378 $ & 0.0  & 31.2  & 14.9  & 43.7  & 0.37 \\
2MASX J15115979-2119015 & 227.9992  & $-21.3171 $ & 7.3  & 11.7  & 7.5  & 43.4  & 0.16 \\
2MASX J15144217-8123377 & 228.6751  & $-81.3939 $ & 5.3  & 12.2  & 7.4  & 43.8  & 0.17 \\
MCG-01-40-001 & 233.3363  & $-8.7005 $ & 0.4  & 6.5  & 3.4  & 42.6  & 0.086 \\
NGC 5995 & 237.1040  & $-13.7578 $ & 0.0  & 2.9  & 2.0  & 42.3  & 0.030 \\
Mrk 1498 & 247.0169  & $51.7754 $ & 1.1  & 6.0  & 3.3  & 43.3  & 0.061 \\
NGC 6240 & 253.2454  & $2.4009 $ & 0.0  & 4.2  & 2.4  & 42.4  & 0.027 \\
NGC 6300 & 259.2478  & $-62.8206 $ & 5.8  & 12.2  & 6.5  & 41.0  & 0.054 \\
2MASX J18074992+1120494 & 271.9580  & $11.3470 $ & 4.7  & 11.6  & 6.0  & \nodata & 0.18 \\
ESO 103-035 & 279.5848  & $-65.4276 $ & 0.3  & 2.9  & 2.3  & 41.7  & 0.011 \\
Fairall 51 & 281.2249  & $-62.3648 $ & 0.5  & 5.9  & 3.7  & 42.1  & 0.060 \\
ESO 141-G055 & 290.3090  & $-58.6703 $ & 0.0  & 2.5  & 1.8  & 42.6  & 0.020 \\
NGC 6814 & 295.6694  & $-10.3235 $ & 0.0\tablenotemark{c} & 6.6  & 3.8  & 41.2  & 0.038 \\
 &  &  & 25.6  & \multicolumn{2}{c}{[Flux] $16\pm5$\tablenotemark{d}, [Index] $\Gamma=2.50\pm0.15$\tablenotemark{d}} & 41.6  & 0.093 \\
NGC 6860 & 302.1954  & $-61.1002 $ & 0.0  & 2.6  & 1.7  & 41.8  & 0.022 \\
Mrk 509 & 311.0406  & $-10.7235 $ & 0.9  & 3.5  & 2.7  & 42.6  & 0.016 \\
6dFJ2132022-334254 & 323.0092  & $-33.7150 $ & 0.0  & 2.1  & 1.3  & 42.3  & 0.020 \\
1RXS J213623.1-622400 & 324.0963  & $-62.4002 $ & 4.3  & 4.8  & 3.6  & 43.2  & 0.071 \\
Mrk 520 & 330.1724  & $10.5524 $ & 0.0  & 3.1  & 2.2  & 42.3  & 0.042 \\
NGC 7172 & 330.5080  & $-31.8698 $ & 2.8  & 4.5  & 3.4  & 41.4  & 0.011 \\
NGC 7213 & 332.3177  & $-47.1667 $ & 4.5  & 6.7  & 3.5  & 40.9  & 0.065 \\
NGC 7314 & 338.9426  & $-26.0503 $ & 2.5  & 6.0  & 3.3  & 40.9  & 0.051 \\
NGC 7319 & 339.0148  & $33.9757 $ & 0.0  & 2.3  & 1.5  & 42.1  & 0.025 \\
Mrk 915 & 339.1938  & $-12.5452 $ & 0.0  & 4.1  & 2.4  & 42.4  & 0.055 \\
MR 2251-178 & 343.5242  & $-17.5819 $ & 0.0  & 2.0  & 1.4  & 42.9  & 0.0085 \\
NGC 7469 & 345.8151  & $8.8740 $ & 0.0  & 2.4  & 1.6  & 41.8  & 0.015 \\
Mrk 926 & 346.1811  & $-8.6857 $ & 0.0  & 2.5  & 1.5  & 42.7  & 0.0094 \\
NGC 7582 & 349.5979  & $-42.3706 $ & 0.3  & 5.8  & 3.5  & 41.0  & 0.031 \\
NGC 7603 & 349.7359  & $0.2440 $ & 0.0  & 3.3  & 2.0  & 42.4  & 0.029 \\
\enddata
\tablenotetext{a}{The case in which the association of ESO~323-G077 with 2FGL~J1306.9--4028 is actually the result of a chance spatial coincidence. 2FGL~J1306.9--4028 is included as a background source in the model.}
\tablenotetext{b}{Assuming that ESO~323-G077 is detected by the LAT as 2FGL~J1306.9--4028.}
\tablenotetext{c}{The case in which the association of NGC~6814 with 2FGL~J1942.5--1024 is actually the result of a chance spatial coincidence. 2FGL~J1942.5--1024 is included as a background source in the model.}
\tablenotetext{d}{Assuming that NGC~6814 is detected by the LAT as 2FGL~J1942.5--1024.}
\tablecomments{(1) Source name from the \bat catalog; 
(2) Right ascension J2000;
(3) Declination J2000;
(4) $TS$ of the $\gamma$-ray event excess using \lat data above 0.2\,GeV at the source position; 
(5) 95\% C.L. upper limits for the photon flux above 0.1\,GeV assuming photon index $\Gamma = 2.5$ ; 
(6) 95\% C.L. upper limits for the photon flux above 0.1\,GeV assuming photon index $\Gamma = 2.2$ ; 
(7) 95\% C.L. upper limits for $\gamma$-ray luminosity in a range between 0.1 and 10\,GeV based on the results of (5), assuming photon index $\Gamma = 2.5$; 
(8) upper limits for $\gamma$-ray--to--hard X-ray luminosity based on the results of (5), assuming photon index $\Gamma = 2.5$; 
}
\end{deluxetable}

\clearpage

\begin{deluxetable}{lrrccccccc}
\tabletypesize{\tiny}
\rotate
\tablecaption{Starburst galaxies discussed in the paper \label{tab:starburst}}
\tablewidth{0pt}
\tablehead{
\colhead{Name} &  \colhead{$d_{\rm L}$} &   \colhead{$\mathcal{F}({\rm >0.1\,GeV})$} &\colhead{$\log{ L_{\gamma}}$}  &  \colhead{$F_{\rm 14-195\,keV}$} &  
\colhead{$\log{ L_{\rm X}}$}  & \colhead{$\log{L_{\rm R}}$} & \colhead{$\log{R_{\rm rX}}$}  & \colhead{$\log{L_{\rm FIR}}$}  & \colhead{$\log{L_{\rm MIR}}$} \\ 
 &  \colhead{[Mpc]} & \colhead{[$10^{-9}$\,ph cm$^{-2}$ s$^{-1}$]}  &  \colhead{[erg s$^{-1}$]} &  \colhead{[$10^{-11}$\,cgs]} &  
 \colhead{[erg s$^{-1}$]}  & \colhead{[erg s$^{-1}$]} &   & \colhead{[erg s$^{-1}$]} & \colhead{[erg s$^{-1}$]}  \\
 \colhead{(1)} &   \colhead{(2)} &  \colhead{(3)} &  \colhead{(4)} &  \colhead{(5)} &  \colhead{(6)} &  \colhead{(7)} &  \colhead{(8)} &  \colhead{(9)} &  \colhead{(10)} 
}
\startdata
NGC 253 & 2.5 & $12.6\pm2.0$ & 39.8  & \nodata & \nodata & 37.77  & \nodata & 43.38  & 42.75  \\
M82 & 3.4 & $15.4\pm1.9$ & 40.2  & \nodata & \nodata & 38.17  & \nodata & 43.83  & \nodata \\
NGC 4945 & 3.7 & $8.5\pm2.8$ & 40.1  & 30.10  & 41.69  & 38.18  & -3.51  & 43.53  & 42.68  \\
NGC 1068 & 16.7 & $6.4\pm2.0$ & 41.2  & 3.38  & 42.05  & 39.37  & -2.68  & 44.35  & 44.44  \\
\enddata
\tablecomments{
(1) source name; 
(2) luminosity distance;
(3) $\gamma$-ray photon flux above 0.1\,GeV taken from \citet{SB};
(4) $\gamma$-ray luminosity above 0.1\,GeV taken from \citet{SB};
(5) $14 -195$\,keV energy flux from the \bat 58-month catalog;
(6) $14 -195$\,keV luminosity;
(7) $1.4$\,GHz radio luminosity;
(8) hard X-ray radio loudness parameter;
(9) FIR luminosity at 90\,\micron\ from the {\it AKARI}--FIS data for NGC~1068, and at 60\,\micron\ from the IRAS data for others; 
(10) MIR luminosity at 9\,\micron\ from the {\it AKARI}--IRC data. 
}
\end{deluxetable}


\begin{thebibliography}{99}

\bibitem[Abdo et al.(2009a)]{Calib}
Abdo, A.~A., et al.\ 2009a, Astroparticle Physics, 32, 193

\bibitem[Abdo et al.(2009b)]{NLS} 
Abdo, A.~A., et al.\ 2009b, \apjl, 707, L142 


\bibitem[Abdo et al.(2010a)]{1LAC}
Abdo, A.~A., et al.\ 2010a, \apj, 715, 429 

\bibitem[Abdo et al.(2010b)]{1FGL}
Abdo, A.~A., et al.\ 2010b, \apjs, 188, 405 

\bibitem[Abdo et al.(2010c)]{MAGN} 
Abdo, A.~A., et al.\ 2010c, \apj, 720, 912 

\bibitem[Abdo et al.(2011)]{2FGL}
Abdo, A.~A., et al.\ 2011, submitted to \apjs, arXiv:1108.1435

\bibitem[Ackermann et al.(2011a)]{2LAC} 
Ackermann, M., et al.\ 2011a, \apj, 743, 171 

\bibitem[Ackermann et al.(2011b)]{SB}
Ackermann, M., et al.\ 2011b, submitted to \apj

\bibitem[Ajello et al.(2008)]{Aje08} 
Ajello, M., et al.\ 2008, \apj, 673, 96 

\bibitem[Antonucci(1993)]{ant93} 
Antonucci, R.\ 1993, \araa, 31, 473 

\bibitem[Atwood et al.(2009)]{LAT}   
Atwood, W., et al. 2009 \apj, 697, 1071

\bibitem[Barthelmy et al.(2005)]{bat} 
Barthelmy, S.~D., et al.\ 2005, \ssr, 120, 143 

\bibitem[Baumgartner et al.(2010)]{Bau10}
Baumgartner, W. H., et al. 2010, submitted to \apjs, \\
{\texttt http://heasarc.gsfc.nasa.gov/docs/swift/results/bs58mon/}

\bibitem[Becker et al.(2003)]{FIRST} 
Becker, R.~H., Helfand, D.~J., White, R.~L., Gregg, M.~D., 
\& Laurent-Muehleisen, S.~A.\ 2003, VizieR Online Data Catalog, 8071, 0 

\bibitem[Beckmann et al.(2009)]{bec09} 
Beckmann, V., et al.\ 2009, \aap, 505, 417 

\bibitem[Bignami et al.(1979)]{big79} 
Bignami, G.~F., Fichtel, C.~E., Hartman, R.~C., \& Thompson, D.~J.\ 1979, \apj, 232, 649 

\bibitem[Brunthaler et al.(2000)]{Bru00} 
Brunthaler, A., et al.\ 2000, \aap, 357, L45 

\bibitem[Burlon et al.(2011)]{bur11} 
Burlon, D., Ajello, M., Greiner, J., Comastri, A., Merloni, A., \& Gehrels, N.\ 2011, \apj, 728, 58 

\bibitem[Cappi et al.(2006)]{cap06} 
Cappi, M., et al.\ 2006, \aap, 446, 459 

\bibitem[Cillis et al.(2004)]{Cil04} 
Cillis, A.~N., Hartman, R.~C., \& Bertsch, D.~L.\ 2004, \apj, 601, 142 

\bibitem[Condon et al.(1996)]{con96} 
Condon, J.~J., Helou, G., Sanders, D.~B., \& Soifer, B.~T.\ 1996, \apjs, 103, 81 

\bibitem[Condon et al.(1998)]{NVSS} 
Condon, J.~J., Cotton, W.~D., Greisen, E.~W., Yin, Q.~F., Perley, R.~A., Taylor, G.~B., 
\& Broderick, J.~J.\ 1998, \aj, 115, 1693 

\bibitem[Corbett et al.(2002)]{cor02} 
Corbett, E.~A., et al.\ 2002, \apj, 564, 650 

\bibitem[Corbett et al.(2003)]{cor03} 
Corbett, E.~A., et al.\ 2003, \apj, 583, 670 

\bibitem[Cusumano et al.(2010)]{Cus10} 
Cusumano, G., et al.\ 2010, \aap, 524, A64 

\bibitem[Foschini et al.(2009)]{fos09} 
Foschini, L., Maraschi, L., Tavecchio, F., Ghisellini, G., Gliozzi, M., \& Sambruna, R.~M.\ 2009, Advances in Space Research, 43, 889 

\bibitem[Foschini et al.(2011a)]{fos11a} 
Foschini L., Colpi M., Gallo L., et al. (eds), 2011a, Narrow-Line Seyfert 1 Galaxies and Their Place in the Universe, Proceedings of Science, vol. NLS1, (Trieste: PoS/SISSA), {\texttt http://pos.sissa.it/cgi-bin/reader/conf.cgi?confid=126} 

\bibitem[Foschini et al.(2011b)]{fos11b} 
Foschini, L., Ghisellini, G., Kovalev, Y.~Y., et al.\ 2011b, \mnras, 413, 1671 

\bibitem[Gallimore et al.(2006)]{gal06} 
Gallimore, J.~F., Axon, D.~J., O'Dea, C.~P., Baum, S.~A., \& Pedlar, A.\ 2006, \aj, 132, 546 

\bibitem[Gilli et al.(2007)]{gil07} 
Gilli, R., Comastri, A., \& Hasinger, G.\ 2007, \aap, 463, 79 


\bibitem[Gondek et al.(1996)]{gon96} 
Gondek, D., Zdziarski, A.~A., Johnson, W.~N., George, I.~M., McNaron-Brown, K., Magdziarz, P., Smith, D., \& Gruber, D.~E.\ 1996, \mnras, 282, 646 

\bibitem[Healey et al.(2007)]{CRATES} 
Healey, S.~E., Romani, R.~W., Taylor, G.~B., Sadler, E.~M., Ricci, R., Murphy, T., Ulvestad, 
J.~S., \& Winn, J.~N.\ 2007, \apjs, 171, 61 

\bibitem[Helene(1983)]{hel83} 
Helene, O.\ 1983, Nuclear Instruments and Methods in Physics Research, 212, 319 

\bibitem[Ho(2002)]{ho02} 
Ho, L.~C.\ 2002, \apj, 564, 120 

\bibitem[Ho(2008)]{ho08} 
Ho, L.~C.\ 2008, \araa, 46, 475 

\bibitem[Ho \& Peng(2001)]{ho01} 
Ho, L.~C., \& Peng, C.~Y.\ 2001, \apj, 555, 650 

\bibitem[Ho \& Ulvestad(2001)]{hoR01} 
Ho, L.~C., \& Ulvestad, J.~S.\ 2001, \apjs, 133, 77 


\bibitem[Ho et al.(2001)]{hoX01} 
Ho, L.~C., et al.\ 2001, \apjl, 549, L51 

\bibitem[Inoue et al.(2008)]{Ino08}
Inoue, Y., Totani, T., \& Ueda, Y.\ 2008, \apjl, 672, L5 

\bibitem[Ishihara et al.(2010)]{AKARI-IRC} 
Ishihara, D., et al.\ 2010, \aap, 514, A1

\bibitem[Johnson et al.(1997)]{joh97} 
Johnson, W.~N., McNaron-Brown, K., Kurfess, J.~D., Zdziarski, A.~A., Magdziarz, P., \& Gehrels, N.\ 1997, \apj, 482, 173 

\bibitem[Kataoka et al.(2011)]{kat11} 
Kataoka, J., et al.\ 2011, \apj, 740, 29

\bibitem[Kellermann et al.(1989)]{Kel89} 
Kellermann, K.~I., et al.\ 1989, \aj, 98, 1195

\bibitem[Kukula et al.(1995)]{kuk95} 
Kukula, M.~J., Pedlar, A., Baum, S.~A., \& O'Dea, C.~P.\ 1995, \mnras, 276, 1262 



\bibitem[Lal et al.(2011)]{lal11} 
Lal, D.~V., Shastri, P., \& Gabuzda, D.~C.\ 2011, \apj, 731, 68 

\bibitem[Lenain et al.(2010)]{len10} 
Lenain, J.-P., Ricci, C., T{\"u}rler, M., Dorner, D., \& Walter, R.\ 2010, \aap, 524, A72 

\bibitem[Lin et al.(1993)]{Lin93} 
Lin, Y.~C., et al.\ 1993, \apjl, 416, L53 

\bibitem[Lodato \& Bertin(2003)]{lod03} 
Lodato, G., \& Bertin, G.\ 2003, \aap, 398, 517 

\bibitem[Lubi{\'n}ski et al.(2010)]{lub10} 
Lubi{\'n}ski, P., Zdziarski, A.~A., Walter, R., Paltani, S., Beckmann, V., Soldi, S., Ferrigno, C., \& Courvoisier, T.~J.-L.\ 2010, \mnras, 408, 1851 

\bibitem[Mahadevan et al.(1997)]{mah97} 
Mahadevan, R., Narayan, R., \& Krolik, J.\ 1997, \apj, 486, 268 

\bibitem[Maisack et al.(1995)]{mai95} 
Maisack, M., et al.\ 1995, \aap, 298, 400 

\bibitem[Maisack et al.(1997)]{mai97} 
Maisack, M., Mannheim, K., \& Collmar, W.\ 1997, \aap, 319, 397 


\bibitem[Massaro et al.(2009)]{BZCAT} 
Massaro, E., et al.\ 2009, \aap, 495, 691

\bibitem[Mattox et al.(1996)]{ML} 
Mattox, J.~R., et al.\ 1996, \apj, 461, 396 

\bibitem[Mauch et al.(2008)]{SUMSS} 
Mauch, T., Murphy, T., Buttery, H.~J., Curran, J., Hunstead, R.~W., Piestrzynski, B., Ropbertson, 
J.~G., \& Sadler, E.~M.\ 2008, VizieR Online Data Catalog, 8081, 0

\bibitem[McConville et al.(2011)]{mcc11} 
McConville, W., et al.\ 2011, \apj, 738, 148

\bibitem[Middelberg et al.(2004)]{mid04} 
Middelberg, E., et al.\ 2004, \aap, 417, 925 

\bibitem[Miller et al.(1993)]{Mil93} 
Miller, P., Rawlings, S., \& Saunders, R.\ 1993, \mnras, 263, 425 

\bibitem[Mundell et al.(2003)]{mun03} 
Mundell, C.~G., Wrobel, J.~M., Pedlar, A., \& Gallimore, J.~F.\ 2003, \apj, 583, 192 

\bibitem[Niedzwiecki et al.(2009)]{nie09} 
Niedzwiecki, A., Xie, F.~G., \& Zdziarski, A.~A.\ 2009, The Extreme Sky: Sampling the Universe above 10 keV,  

\bibitem[Oka \& Manmoto(2003)]{oka03} 
Oka, K., \& Manmoto, T.\ 2003, \mnras, 340, 543 

\bibitem[Osterbrock(1977)]{ost77} 
Osterbrock, D.~E.\ 1977, \apj, 215, 733 

\bibitem[Osterbrock(1989)]{ost89} 
Osterbrock, D.~E.\ 1989, Astrophysics of Gaseous Nebulae and Active Galactic Nuclei, (Mill Valley: University Science Books) 

\bibitem[Panessa et al.(2007)]{pan07} 
Panessa, F., Barcons, X., Bassani, L., Cappi, M., Carrera, F.~J., Ho, L.~C., \& Pellegrini, S.\ 2007, \aap, 467, 519 

\bibitem[Pogge(2000)]{pog00} 
Pogge, R.~W.\ 2000, \nar, 44, 381 

\bibitem[Pollock et al.(1981)]{pol81} 
Pollock, A.~M.~T., Masnou, J.~L., Bignami, G.~F., Hermsen, W., Swanenburg, B.~N., Kanbach, G., Lichti, G.~G., \& Wills, R.~D.\ 1981, \aap, 94, 116 

\bibitem[Poutanen(1998)]{pou98} 
Poutanen, J.\ 1998, Theory of Black Hole Accretion Disks, 100 


\bibitem[Sanders et al.(2003)]{IRAS} 
Sanders, D.~B., Mazzarella, J.~M., Kim, D.-C., Surace, J.~A., \& Soifer, B.~T.\ 2003, \aj, 126, 1607 

\bibitem[Schoenfelder et al.(1993)]{comptel} 
Schoenfelder, V., et al.\ 1993, \apjs, 86, 657 

\bibitem[Seyfert(1943)]{Sey43} 
Seyfert, C.~K.\ 1943, \apj, 97, 28 

\bibitem[Shapiro et al.(1976)]{shap76} 
Shapiro, S.~L., Lightman, A.~P., \& Eardley, D.~M.\ 1976, \apj, 204, 187

\bibitem[Sikora et al.(2007)]{sik07} 
Sikora, M., Stawarz, {\L}., \& Lasota, J.~P.\ 2007, \apj, 658, 815 

\bibitem[Svensson(1987)]{sve87} 
Svensson, R.\ 1987, \mnras, 227, 403 

\bibitem[Teng et al.(2011)]{ten11} 
Teng, S.~H., Mushotzky, R.~F., Sambruna, R.~M., Davis, D.~S., 
\& Reynolds, C.~S.\ 2011, \apj, 742, 66 

\bibitem[Terashima \& Wilson(2003)]{Ter03} 
Terashima, Y., \& Wilson, A.~S.\ 2003, \apj, 583, 145 

\bibitem[Thompson et al.(1993)]{egret} 
Thompson, D.~J., et al.\ 1993, \apjs, 86, 629 

\bibitem[Tueller et al.(2008)]{tue08} 
Tueller, J., Mushotzky, R.~F., Barthelmy, S., Cannizzo, J.~K., Gehrels, N., Markwardt, C.~B., Skinner, G.~K., \& Winter, L.~M.\ 2008, \apj, 681, 113 

\bibitem[Ulvestad \& Wilson(1984)]{ulv84} 
Ulvestad, J.~S., \& Wilson, A.~S.\ 1984, \apj, 285, 439

\bibitem[Ulvestad \& Wilson(1989)]{ulv89} 
Ulvestad, J.~S., \& Wilson, A.~S.\ 1989, \apj, 343, 659 

\bibitem[Ulvestad \& Ho(2001)]{ulv01}
Ulvestad, J.~S., \& Ho, L.~C.\ 2001, \apj, 558, 561

\bibitem[Ulvestad et al.(2005)]{ulv05} 
Ulvestad, J.~S., Wong, D.~S., Taylor, G.~B., Gallimore, J.~F., \& Mundell, C.~G.\ 2005, \aj, 130, 936 


\bibitem[Wardzi{\'n}ski \& Zdziarski(2001)]{war01} 
Wardzi{\'n}ski, G., \& Zdziarski, A.~A.\ 2001, \mnras, 325, 963 

\bibitem[White \& Becker(1992)]{whi92} 
White, R.~L., \& Becker, R.~H.\ 1992, \apjs, 79, 331 

\bibitem[Wright \& Otrupcek(1990)]{PKS} 
Wright, A., \& Otrupcek, R.\ 1990, PKS Catalog (1990), 0

\bibitem[Yamamura et al.(2010)]{AKARI-FIS} 
Yamamura, I., Makiuti, S., Ikeda, N., Fukuda, Y., Oyabu, S., Koga, T., 
\& White, G.~J.\ 2010, VizieR Online Data Catalog, 2298, 0 

\bibitem[Yun et al.(2010)]{yun01} 
Yun, M. S., Reddy, N. A., \& Condon, J. J. 2001, \apj, 554, 803

\bibitem[Zdziarski(1999)]{zdz99} 
Zdziarski, A.~A.\ 1999, High Energy Processes in Accreting Black Holes, 161, 16 

\bibitem[Zdziarski \& Lightman(1985)]{zdz85} 
Zdziarski, A.~A., \& Lightman, A.~P.\ 1985, \apjl, 294, L79 

\bibitem[Zdziarski et al.(2000)]{zdz00} 
Zdziarski, A.~A., Poutanen, J., \& Johnson, W.~N.\ 2000, \apj, 542, 703 

\bibitem[Zhou \& Zhang(2010)]{zho10} 
Zhou, X.-L., \& Zhang, S.-N.\ 2010, \apjl, 713, L11 

      
\end{thebibliography}
\end{document}